\expandafter\ifx\csname qd\endcsname\relax
   \documentstyle[12pt,twoside]{article}\let\qd=\quad \let\qqd=\qquad \def\qqqd{\qquad\qquad}

\let\a=\alpha \let\be=\beta \let\g=\gamma \let\de=\delta
\let\e=\varepsilon \let\z=\zeta \let\h=\eta \let\th=\theta
\let\dh=\vartheta \let\k=\kappa \let\la=\lambda
\let\r=\rho \let\s=\sigma
\let\om=\omega
\let\ph=\varphi
\let\Om=\Omega \let\Si=\Sigma \let\Th=\Theta
\let\La=\Lambda \let\G=\Gamma \let\D=\Delta

\def\0{\over } \def\1{\vec } \def\2{{1\over2}} \def\4{{1\over4}}
\def\5{\bar } \def\6{\partial } \def\7#1{{#1}\llap{/}}
\def\8{\tilde } \def\9{\dot }

\def\ul#1{$\underline{\hbox{#1}}$} \def\llp{\hbox to 0pt{\hss /\hskip1.5pt}}
\def\so{\supset\hbox to 0pt{\hss $\displaystyle -$\hskip1pt}}

\def\ts{\textstyle } \def\ds{\displaystyle }
\def\<{\langle } \def\>{\rangle } \def\lb{\left} \def\rb{\right}
\let\lra=\leftrightarrow \let\LRA=\Leftrightarrow
\let\then=\Rightarrow \def\triang{{\scriptstyle \Delta}}
\def\trian{{\scriptscriptstyle \Delta}}
\def\re{{\rm Re}} \def\im{{\rm Im}} \def\ord{{\cal O}}

\let\ap=\approx \let\eq=\equiv \let\ex=\times \let\hc=\dagger
      \let\ti=\tilde \let\bl=\biggl \let\br=\biggr
\let\bi=\choose \let\at=\atop \let\mat=\pmatrix

\let\nn=\nonumber
\def\bea{\begin{eqnarray}} \def\eea{\end{eqnarray}}
\def\bean{\begin{eqnarray*}} \def\eean{\end{eqnarray*}}
\def\beq{\begin{equation}} \def\eeq{\end{equation}}
\def\barr{\begin{array}} \def\earr{\end{array}}

\newcommand{\deflabel}[1]{#1\hfil}%
{\begin{list}{}%
{\settowidth{\labelwidth}{#1}%
\setlength{\leftmargin}{\labelwidth}\addtolength{\leftmargin}{\labelsep}%
\renewcommand{\makelabel}{\deflabel}}}%
{\end{list}}%
\fi

\begin{document}
\noindent
{\pagestyle{empty}
June 1994 \\
\vspace{1cm}
\begin{center}
 {\Large\bf The semiclassical stability\\
              of de~Sitter spacetime}\\
 \vspace{2cm}
 C. Busch \bigskip\\
 Institut f\"ur Theoretische Physik, Universit\"at Hannover \\
 Appelstr. 2, D-30167 Hannover, Germany \\
 \vspace{2cm}
 {\large\bf Abstract}\bigskip\\
 \parbox{\textwidth}{\small
  de~Sitter spacetime and Bunch-Davies vacuum are a solution to the
  semiclassical Einstein-Schr\"odinger equations describing the evolution
  of spacetime geometry and a massive scalar quantum field with arbitrary
  coupling to curvature. The stability of this solution is proven by
  calculating the renormalized energy momentum tensor expectation value
  for small spatially homogeneous deviations from the
  de~Sitter -- Bunch-Davies system and solving the linearized backreaction
  problem. A renormalization scheme is developed. All momentum integrations
  are carried out analytically. The general solution is given in terms of
  its Laplace transform. It contains only two artificial instabilities:
  a constant gauge mode and an instability on the Planck
  time scale lying outside of the scope of our semiclassical theory.}
\end{center}
\vfill
\pageref{LastPage} pages, 3 figures
}
\clearpage
\expandafter\ifx\csname qd\endcsname\relax
   \documentstyle[12pt,twoside]{article}\let\qd=\quad \let\qqd=\qquad \def\qqqd{\qquad\qquad}

\let\a=\alpha \let\be=\beta \let\g=\gamma \let\de=\delta
\let\e=\varepsilon \let\z=\zeta \let\h=\eta \let\th=\theta
\let\dh=\vartheta \let\k=\kappa \let\la=\lambda
\let\r=\rho \let\s=\sigma
\let\om=\omega
\let\ph=\varphi
\let\Om=\Omega \let\Si=\Sigma \let\Th=\Theta
\let\La=\Lambda \let\G=\Gamma \let\D=\Delta

\def\0{\over } \def\1{\vec } \def\2{{1\over2}} \def\4{{1\over4}}
\def\5{\bar } \def\6{\partial } \def\7#1{{#1}\llap{/}}
\def\8{\tilde } \def\9{\dot }

\def\ul#1{$\underline{\hbox{#1}}$} \def\llp{\hbox to 0pt{\hss /\hskip1.5pt}}
\def\so{\supset\hbox to 0pt{\hss $\displaystyle -$\hskip1pt}}

\def\ts{\textstyle } \def\ds{\displaystyle }
\def\<{\langle } \def\>{\rangle } \def\lb{\left} \def\rb{\right}
\let\lra=\leftrightarrow \let\LRA=\Leftrightarrow
\let\then=\Rightarrow \def\triang{{\scriptstyle \Delta}}
\def\trian{{\scriptscriptstyle \Delta}}
\def\re{{\rm Re}} \def\im{{\rm Im}} \def\ord{{\cal O}}

\let\ap=\approx \let\eq=\equiv \let\ex=\times \let\hc=\dagger
      \let\ti=\tilde \let\bl=\biggl \let\br=\biggr
\let\bi=\choose \let\at=\atop \let\mat=\pmatrix

\let\nn=\nonumber
\def\bea{\begin{eqnarray}} \def\eea{\end{eqnarray}}
\def\bean{\begin{eqnarray*}} \def\eean{\end{eqnarray*}}
\def\beq{\begin{equation}} \def\eeq{\end{equation}}
\def\barr{\begin{array}} \def\earr{\end{array}}

\newcommand{\deflabel}[1]{#1\hfil}%
\newenvironment{deflist}[1]%
{\begin{list}{}%
{\settowidth{\labelwidth}{#1}%
\setlength{\leftmargin}{\labelwidth}\addtolength{\leftmargin}{\labelsep}%
\renewcommand{\makelabel}{\deflabel}}}%
{\end{list}}%
\fi
\expandafter\ifx\csname ok\endcsname\relax
   \def\re{{\rm Re}} \def\im{{\rm Im}}
   {\catcode`@=11 \@addtoreset{equation}{section}\@addtoreset{figure}{section}}
   \def\theequation{\thesection .\arabic{equation}}
   \def\thefigure{\thesection .\arabic{figure}}
   \addtolength{\topmargin}{-48pt}
   \addtolength{\textheight}{90pt}
   \addtolength{\evensidemargin}{-31pt}
   \addtolength{\oddsidemargin}{-12pt}
   \addtolength{\textwidth}{30pt}
   \addtolength{\footskip}{21pt}
   \setlength{\parindent}{0pt}\frenchspacing
   \begin{document}
\fi
\section{Introduction}
The de~Sitter spacetime \cite{deSitter} is of great theoretical as well as
cosmological interest. The former arises due to its high degree of
symmetry: with 10 Killing vector fields its isometry group O(4,1) has the same
dimension as the Poincar\'e group of Minkowski spacetime and therefore the
maximum dimension the symmetry group of a four dimensional spacetime can have
at all. Just this fact makes a lot of calculations of quantum field theory
feasible in the de~Sitter spacetime.\\
The cosmological interest stems from the exponential growth of the scale factor
in the spatially flat (k=0) Friedmann-Robertson-Walker parametrization of
de~Sitter spacetime solving some basic problems of the standard cosmology
in inflationary universe models \cite{Inflation}.\\
In absence of a consistent quantum theory of gravity one usually works within
a semiclassical framework, where the gravitational field is treated as a
classical background field and only the matter fields are quantized. This is
justified as long as all relevant inverse time and length scales are small
compared to the Planck scale\footnote{in natural units $\hbar=c=1$},
so that quantum gravity effects are expected to be small.\\
Since the work of Schwinger it is known that the quantum fluctuations of a
charged matter field in an electromagnetic background field can lead to the
production of particle-antiparticle pairs. The same applies to the
gravitational background field and is known as the Hawking effect.
From this observation the conjecture and also some claims arose in the
literature (see for example \cite{Lit1}-\cite{Mottola}), that in the presence
of a scalar quantum field like in most inflationary
scenarios the de~Sitter spacetime might be unstable due to particle production
and should decay in some sense by itself towards a flat spacetime.\\
There is no unique observer-independent particle-antiparticle concept in a
general curved spacetime and different approaches to particle production
involving diffe\-rent approximations led to different answers for this stability
question.\\
In reference \cite{TraHill} on the contrary this question is adressed in a
rather clear and reliable manner (see below) based on the energy momentum
tensor, an observer-independent physical quantity. Unfortunately no sensible
results were obtained due to technical problems, on which we will comment later.
In a further publication \cite{GuvLieb} the main problems were
not eliminated. Nevertheless their approach is promising and will be adopted
as the starting point for the present investigation.\\
Within the semiclassical theory the evolution of spacetime geometry is
governed by the Einstein equations containing the expectation value of the
energy momentum tensor as the source on the right-hand side, whereas the
quantum state of the scalar field has to obey the Schr\"odinger equation, which
in turn depends on the spacetime metric:\\
\parbox{\textwidth}{
\bea R_{\mu\nu}-\2\,g_{\mu\nu}R+\La\,g_{\mu\nu}&=&8\pi G_N\,
  \<\Psi|T_{\mu\nu}|\Psi\>\nn\\
  i\,\6_t\,|\Psi\>&=&\hat{H}_g\,|\Psi\> \label{EinSchro}\eea}
The de~Sitter spacetime and the Bunch-Davies vacuum, a special state of the quantum
field, are a solution to this semiclassical system of coupled equations. In
order to investigate the stability of this solution against small fluctuations
of the gravitational field and of the quantum state we will linearize the
equations (\ref{EinSchro}) around the de~Sitter -- Bunch-Davies solution.
This linearization is the only
approximation appearing within this work.
The linearized Einstein-Schr\"odinger equations will be solved completely
and the general solution will be analyzed with respect to instabilities.\\
In the course of this work another publication \cite{IsaRog} on the same subject
appeared also based on reference \cite{TraHill}. We will
reach the same conclusions as reference \cite{IsaRog} but in a more direct
way, because in the calculation of the energy momentum tensor we will execute
the momentum integrations analytically, so that the result is suited for a
numerical analysis. Furthermore a general coupling of the scalar quantum field
to curvature will be allowed.\\
The paper is organized as follows:
section \ref{FRWQFT} gives a brief review of some important
results from reference \cite{QFT}, whereas section \ref{Fast}, the main part
of this work, contains the calculation of the energy momentum tensor
expectation value and the isolation of its divergencies. The linear
stability analysis is performed in section \ref{solv}. Some mathematical tools
are collected in the appendix.

\expandafter\ifx\csname ok\endcsname\relax
   \end{document} \fi
\expandafter\ifx\csname qd\endcsname\relax
   \documentstyle[12pt,twoside]{article}\let\qd=\quad \let\qqd=\qquad \def\qqqd{\qquad\qquad}

\let\a=\alpha \let\be=\beta \let\g=\gamma \let\de=\delta
\let\e=\varepsilon \let\z=\zeta \let\h=\eta \let\th=\theta
\let\dh=\vartheta \let\k=\kappa \let\la=\lambda
\let\r=\rho \let\s=\sigma
\let\om=\omega
\let\ph=\varphi
\let\Om=\Omega \let\Si=\Sigma \let\Th=\Theta
\let\La=\Lambda \let\G=\Gamma \let\D=\Delta

\def\0{\over } \def\1{\vec } \def\2{{1\over2}} \def\4{{1\over4}}
\def\5{\bar } \def\6{\partial } \def\7#1{{#1}\llap{/}}
\def\8{\tilde } \def\9{\dot }

\def\ul#1{$\underline{\hbox{#1}}$} \def\llp{\hbox to 0pt{\hss /\hskip1.5pt}}
\def\so{\supset\hbox to 0pt{\hss $\displaystyle -$\hskip1pt}}

\def\ts{\textstyle } \def\ds{\displaystyle }
\def\<{\langle } \def\>{\rangle } \def\lb{\left} \def\rb{\right}
\let\lra=\leftrightarrow \let\LRA=\Leftrightarrow
\let\then=\Rightarrow \def\triang{{\scriptstyle \Delta}}
\def\trian{{\scriptscriptstyle \Delta}}
\def\re{{\rm Re}} \def\im{{\rm Im}} \def\ord{{\cal O}}

\let\ap=\approx \let\eq=\equiv \let\ex=\times \let\hc=\dagger
      \let\ti=\tilde \let\bl=\biggl \let\br=\biggr
\let\bi=\choose \let\at=\atop \let\mat=\pmatrix

\let\nn=\nonumber
\def\bea{\begin{eqnarray}} \def\eea{\end{eqnarray}}
\def\bean{\begin{eqnarray*}} \def\eean{\end{eqnarray*}}
\def\beq{\begin{equation}} \def\eeq{\end{equation}}
\def\barr{\begin{array}} \def\earr{\end{array}}

\newcommand{\deflabel}[1]{#1\hfil}%
\newenvironment{deflist}[1]%
{\begin{list}{}%
{\settowidth{\labelwidth}{#1}%
\setlength{\leftmargin}{\labelwidth}\addtolength{\leftmargin}{\labelsep}%
\renewcommand{\makelabel}{\deflabel}}}%
{\end{list}}%
\fi
\expandafter\ifx\csname ok\endcsname\relax
   \def\re{{\rm Re}} \def\im{{\rm Im}}
   {\catcode`@=11 \@addtoreset{equation}{section}\@addtoreset{figure}{section}}
   \def\theequation{\thesection .\arabic{equation}}
   \def\thefigure{\thesection .\arabic{figure}}
   \addtolength{\topmargin}{-48pt}
   \addtolength{\textheight}{90pt}
   \addtolength{\evensidemargin}{-31pt}
   \addtolength{\oddsidemargin}{-12pt}
   \addtolength{\textwidth}{30pt}
   \addtolength{\footskip}{21pt}
   \setlength{\parindent}{0pt}\frenchspacing
   \begin{document}
\fi
\section[Schr\"odinger picture field theory in k=0
 Friedmann-Robertson-Walker\protect\\ spacetimes]
{Schr\"odinger picture field theory in k=0\protect\\
 Friedmann-Robertson-Walker spacetimes}
\label{FRWQFT}
This section deals with the quantum theory of a free, massive scalar field in
a spatially flat (k=0) Friedmann-Robertson-Walker (FRW) spacetime, and it is
a brief summary of some results from reference \cite{QFT}.\\
Since dimensional regularization will be applied later on, we work in $d{+}1$
spacetime dimensions and on flat $d$-dimensional spacelike hyperplanes with
coordinates $\1x=(x_1,\ldots,x_d)$. The maximum symmetry of these hyperplanes
will greatly simplify the following calculations. The k=0 FRW-metric has the
form
\beq ds^2 = dt^2 - a^2(t)\,d\vec{x}\cdot d\vec{x} = g_{\mu\nu}dx^{\mu}dx^{\nu},
     \label{Met}\eeq
where $a(t)$ is the FRW scale factor. In terms of the Hubble function
$H(t):=\9a(t)/a(t)$, $\9a:=\6_ta$, one obtains for the Ricci tensor
\beq R_{00}=-d(\dot{H}+H^2)\;,\qqd R_{ij}=a^2(\dot{H}+dH^2)\de_{ij}\;,\qqd
     R_{0i}=0 \eeq
and for the curvature scalar $R=-d(2\dot{H}+(d+1)H^2)$.\\
A scalar field $\phi$ of mass $m$ is supposed to interact only with the
classical gravitational field $g_{\mu\nu}$ and may have an arbitrary coupling
$\xi$ to the curvature scalar $R$. Its action reads ($\sqrt{g}:=a^d$)
\beq S=\int\!d^{d+1}\!x\,\sqrt{g}\,\2\,(g^{\mu\nu}(\6_{\mu}\phi)(\6_{\nu}\phi)
       -(m^2\!-\!\xi R)\phi^2). \label{Wirk}\eeq
The quantum theory is formulated in the Schr\"odinger picture using a wave
functional to represent the quantum state. This shows very clearly
the real time evolution character of our analysis. Then the quantum operators
are acting in the Fock space on wave functionals $\Psi[\phi;t]$.\\
The spacelike derivatives contained in the Hamiltonian can be dealt with by
performing a Fourier transform ($d\8k:=d^d\!k/(2\pi)^{d\!/\!2}$,
 $\a^*(\1k)=\a(-\1k)$):
\beq \phi(\1x)\,=\,\int d\8k\,e^{i\1k\1x}\,\a(\1k)\;,\qqd
     {\de\0\de\phi(\1x)}\,=\,\int d\8k\,e^{-i\1k\1x}\,{\de\0\de\a(\1k)}\;,
     \label{Fourier}\eeq
and the Schr\"odinger equation resp. the Hamiltonian takes the following form:
\bea i\,\6_t\,\Psi[\a;t]\!&=&\!\hat{H}\,\Psi[\a;t] \label{Schroe}\\
     \hat{H}\!&=&\!\2\int d^d\!k\,\Bigl(-{1\0\sqrt{g}}
     {\de^2\0\de\a(\1k)\de\a(-\1k)}+\sqrt{g}(a^{-2}\1k^2\!+\!m^2\!-\!\xi R)
     \a(\1k)\a(-\1k)\Bigr) \nn\eea
For a curved
spacetime without an everywhere timelike Killing vector field no unique Fock
vacuum does exist. Rather there is a whole class of Fock vacua, which can all
be represented by Gaussian wave functionals. We want our $\Psi[\a;t]$ to be
a member of this class. Furthermore we require the quantum state
not to break spontaneously the symmetries of the k=0 FRW metric (homogeneity
and isotropy of the spacelike hyperplanes), which leads to the following
wave functional parametrized by one function $A(k,t)$ (the inverse Gaussian
width):
\beq \Psi[\a;t]\;=\;N(t)\exp\Bigl(-\2\int d^d\!k\,A(k,t)\,\a(\1k)\,\a(-\1k)-i\Om(t)
     \Bigr), \label{Gaus}\eeq
where $N(t)$ is a real normalization factor, $\Om(t)$ a real phase and
$k:=|\1k|$.\\
Substituting (\ref{Gaus}) in the Schr\"odinger equation (\ref{Schroe}) we get the
equation of motion for $A(k,t)$:
\beq i\,\9A(k,t)\;=\;{A^2(k,t)\0\sqrt{g(t)}}
     -\sqrt{g(t)}\,(a^{-2}(t)k^2\!+\!m^2\!-\!\xi R(t))  \label{Bew1}\eeq
This is Riccati's equation, and by the transformation
$A(k,t)=:\sqrt{g}(\G(k,t)+i{d\02}H(t))$ it takes the standard form
\beq i\,\9{\G}(k,t)\;=\;\G^2(k,t)+{d^2\04}H^2(t)+{d\02}\9H(t)
      -(a^{-2}(t)k^2\!+\!m^2\!-\!\xi R(t)), \label{Bew2}\eeq
which can be converted by $\G(k,t)=:-i\6_t\ln u(k,t)$ into the linear equation
\beq \ddot{u}\;-\;({d^2\04}H^2+{d\02}\9H-(a^{-2}k^2\!+\!m^2\!-\!\xi R))\,u\;=\;0
  \qd. \label{Bew3}\eeq
From (\ref{Bew1}) the following equations for $A(k,t)$ can be derived, which
are useful for the calculation of the energy momentum expectation value:
\beq -\6_t\,\lb({1\02\,\re A}\rb)\;=\;{1\0\sqrt{g}}\,{\im A\0\re A}
     \label{Bew4}\eeq
\beq {1\0g}\,{|A|^2\02\,\re A}\;=\;(a^{-2}k^2\!+\!m^2\!-\!\xi R)\,{1\02\,\re A}
 +{1\02\,\sqrt{g}}\,\6_t\lb(\sqrt{g}\6_t\lb({1\02\,\re A}\rb)\rb)\label{Bew5}\eeq
The energy momentum tensor acting as the source in the Einstein equations
is defined as variational derivative of the matter action with respect to the
metric tensor
\[ T_{\mu\nu}(x)\;:=\;{2\0\sqrt{g}}\,\,{\de S\0\de g^{\mu\nu}(x)}\;.\]
Due to the spatial symmetries the expectation value of the corresponding
operator in the Gaussian state can be written as
\beq \<\Psi|({T^{\mu}}_{\nu})|\Psi\>\;=\;\left(\begin{array}[c]{cccc}
   \r(t)\\&-p(t)\\&&-p(t)\\&&&-p(t)\end{array}\right)\;,\eeq
and the explicit calculation leads to the energy density $\r$ and pressure $p$
in terms of the width $A(k,t)$:
\bea \r\!&=&\!
   \2\int{\8d\8k\02\,\re A(k,t)}\biggl({|A(k,t)|^2\0g}+a^{-2}k^2+m^2
   +2\xi\, G_{00}-2\xi\,{dH\0\sqrt{g}}\,2\,\im A(k,t)\biggr) \nn\\
   p\!&=&\!
   \2\int{\8d\8k\02\,\re A(k,t)}\biggl({|A(k,t)|^2\0g}
   +({2\0d}\!-\!1)\,a^{-2}k^2-m^2+2\xi\, a^{-2}G_{11} \nn\\
  &&\!\qd+\;4\xi\,(a^{-2}k^2\!+\!m^2\!-\!\xi R)-4\xi\,{|A(k,t)|^2\0g}-2\xi\,
   {H\0\sqrt{g}}\,2\,\im A(k,t)\biggr) \label{EIT4}\eea
$G_{\mu\nu}=R_{\mu\nu}-\2\,g_{\mu\nu}\,R$ is the Einstein tensor,
and $\8d\8k:=d^d\!k/(2\pi)^d$.\\
Equation (\ref{Bew1}) shows, that $A(k,t)$ is of the order of $k$ for large $k$.
Hence the energy momentum expectation value (\ref{EIT4}) is quartic divergent
and has to be renormalized. These ultraviolet divergencies are due to the
behaviour of the wave functional for field configurations of high energy and
momenta (large $k$) or resp. for small distances and are connected to the local
geometry of the underlying spacetime manifold. For this reason they should be
proportional to local geometric tensors which can be absorbed into the
gravitational part of the Einstein equations. Thus the divergencies can be
removed by a renormalization of the physical parameters in the Einstein
equations (cosmological constant, Newton's constant and additional parameters
mentioned below). Fortunately the divergencies of the energy momentum
expectation value can be calculated as a local functional of a general metric
tensor by means of the De~Witt-Schwinger-Christensen expansion.
It turns out that in the Einstein equations one has to admit
the geometrical tensors $H_{\mu\nu}$, $^{(1)}\!H_{\mu\nu}$ and
$^{(2)}\!H_{\mu\nu}$, which are the metric variations
$1/\sqrt{g}\;\de/\de g^{\mu\nu}$ of the functionals
$\int d^{d\!+\!1}\!x\,\sqrt{g}\,R^{\a\be\r\s}R_{\a\be\r\s}$,
$\int d^{d\!+\!1}\!x\,\sqrt{g}\,R^2$ and
$\int d^{d\!+\!1}\!x\,\sqrt{g}\,R^{\a\be}R_{\a\be}$.
Their renormalized coefficients have to be regarded as additional physical
parameters of the theory.
We will choose them to be zero, since the effects of these terms have already
been analyzed elsewhere \cite{Kuss}.\\
The renormalization scheme consists in a subtraction
of the first three divergent terms of the De~Witt-Schwinger-Christensen series
from the expectation value (\ref{EIT4}):
\beq \<T_{\mu\nu}\>_{\rm ren}\;:=\;
  \<\Psi|T_{\mu\nu}|\Psi\>-\<T_{\mu\nu}\>_{\rm DS\,div} \eeq
If $|\Psi\>$ is a state of finite energy density (compared with an adiabatic
vacuum as will be explained later on), then the divergencies of
$\<\Psi|T_{\mu\nu}|\Psi\>$ and $\<T_{\mu\nu}\>_{\rm DS\,div}$ will cancel
and $\<T_{\mu\nu}\>_{\rm ren}$ is finite. It should be noted, that the
renormalization scheme decides about the physical meaning of the renormalized
parameters.\\
With dimensional regularization ($d=3-\e$) one obtains \cite{BirDav}\\
\parbox{\textwidth}{
\bea \<T_{\mu\nu}\>_{\rm DS\,div}&=&{1\016\pi^2}\,\biggl({1\0\e}-\2\Bigl(\g+
   \ln{m^2\04\pi}\Bigr)\biggr)\!\cdot\!
   \biggl({-4\,m^4\0(d\!+\!1)(d\!-\!1)}\,g_{\mu\nu}-{4\,m^2\0d-1}
   (\xi\!-\!{1\06})G_{\mu\nu}\nn\\
   & &\qqqd\qqqd+\;{1\090}(H_{\mu\nu}-{}^{(2)}\!H_{\mu\nu})+
   (\xi\!-\!{1\06})^2\,{}^{(1)}\!H_{\mu\nu}\biggr),\label{TDS2}\eea}
where $\g$ is the Euler-Mascheroni constant. For our FRW metric (\ref{Met})
the $H$-tensors are explicitly given in appendix \ref{geoT}.
\expandafter\ifx\csname ok\endcsname\relax
   \end{document}\fi

\expandafter\ifx\csname qd\endcsname\relax
   \documentstyle[12pt,twoside]{article}\let\qd=\quad \let\qqd=\qquad \def\qqqd{\qquad\qquad}

\let\a=\alpha \let\be=\beta \let\g=\gamma \let\de=\delta
\let\e=\varepsilon \let\z=\zeta \let\h=\eta \let\th=\theta
\let\dh=\vartheta \let\k=\kappa \let\la=\lambda
\let\r=\rho \let\s=\sigma
\let\om=\omega
\let\ph=\varphi
\let\Om=\Omega \let\Si=\Sigma \let\Th=\Theta
\let\La=\Lambda \let\G=\Gamma \let\D=\Delta

\def\0{\over } \def\1{\vec } \def\2{{1\over2}} \def\4{{1\over4}}
\def\5{\bar } \def\6{\partial } \def\7#1{{#1}\llap{/}}
\def\8{\tilde } \def\9{\dot }

\def\ul#1{$\underline{\hbox{#1}}$} \def\llp{\hbox to 0pt{\hss /\hskip1.5pt}}
\def\so{\supset\hbox to 0pt{\hss $\displaystyle -$\hskip1pt}}

\def\ts{\textstyle } \def\ds{\displaystyle }
\def\<{\langle } \def\>{\rangle } \def\lb{\left} \def\rb{\right}
\let\lra=\leftrightarrow \let\LRA=\Leftrightarrow
\let\then=\Rightarrow \def\triang{{\scriptstyle \Delta}}
\def\trian{{\scriptscriptstyle \Delta}}
\def\re{{\rm Re}} \def\im{{\rm Im}} \def\ord{{\cal O}}

\let\ap=\approx \let\eq=\equiv \let\ex=\times \let\hc=\dagger
      \let\ti=\tilde \let\bl=\biggl \let\br=\biggr
\let\bi=\choose \let\at=\atop \let\mat=\pmatrix

\let\nn=\nonumber
\def\bea{\begin{eqnarray}} \def\eea{\end{eqnarray}}
\def\bean{\begin{eqnarray*}} \def\eean{\end{eqnarray*}}
\def\beq{\begin{equation}} \def\eeq{\end{equation}}
\def\barr{\begin{array}} \def\earr{\end{array}}

\newcommand{\deflabel}[1]{#1\hfil}%
\newenvironment{deflist}[1]%
{\begin{list}{}%
{\settowidth{\labelwidth}{#1}%
\setlength{\leftmargin}{\labelwidth}\addtolength{\leftmargin}{\labelsep}%
\renewcommand{\makelabel}{\deflabel}}}%
{\end{list}}%
\fi
\expandafter\ifx\csname ok\endcsname\relax
   \def\re{{\rm Re}} \def\im{{\rm Im}}
   {\catcode`@=11 \@addtoreset{equation}{section}\@addtoreset{figure}{section}}
   \def\theequation{\thesection.\arabic{equation}}
   \def\thefigure{\thesection.\arabic{figure}}
   \addtolength{\topmargin}{-48pt}
   \addtolength{\textheight}{90pt}
   \addtolength{\evensidemargin}{-31pt}
   \addtolength{\oddsidemargin}{-12pt}
   \addtolength{\textwidth}{30pt}
   \addtolength{\footskip}{21pt}
   \setlength{\parindent}{0pt}\frenchspacing
   \includeonly{}
   \begin{document}
   
\fi
\subsection{de~Sitter spacetime and Bunch-Davies vacuum}\label{DeSit}
As already mentioned in the introduction we use the k=0 FRW parametrization
of the de~Sitter spacetime: $a(t)=e^{H_0t}$ in (\ref{Met}). $H_0$ is the
Hubble constant and due to the maximal symmetry we have
$R_{\mu\nu}=-d\,H_0^2\,g_{\mu\nu}$ and $R=-d(d\!+\!1)\,H_0^2\,$.\\
By substituting the conformal time $\tau(t):=e^{-H_0t}/H_0$ equation
(\ref{Bew3}) becomes a Bessel differential equation:
\beq ((k\tau)^2\,\6_{k\tau}^2+k\tau\,\6_{k\tau}+(k\tau)^2-\nu^2)\,\,
     u(k,t)\;=\;0 \label{Bessel}\eeq
where $\nu^2:=d^2/4-(m^2\!-\!\xi R)/H_0^2$.\\
Its general solution is a linear combination of the two Hankel functions
$H^{(1)}_{\nu}$ and $H^{(2)}_{\nu}$:
\beq u(k,t)\;=\;B_1(k)\,H^{(1)}_{\nu}(k\tau)+B_2(k)\,H^{(2)}_{\nu}(k\tau)
    \label{u1}\eeq
A comparison with the adiabatic vacuum in section \ref{adia} shows that
we have to require $B_1(k)\stackrel{k\to\infty}{\to}0\,$, since our quantum
state should have a finite energy density.\\
In addition we want the quantum state not to break the de~Sitter
symmetry. It follows that $\r$ and $p$ have to be constant over the whole
de~Sitter manifold. This is the case if $B_1$ and $B_2$ in (\ref{u1}) are
independent of $k$, as can be seen by substituting $y:=k\tau$ in the
integrals (\ref{EIT4}).\\
Therefore we end up with $B_1(k)=0$ and $u(k,t)=H^{(2)}_{\nu}(k\tau(t))$,
leading to
\beq {1\02\,\re A(k,t)}\;=\;{\pi\04}\,H_0^{d-1}\tau^d
     H^{(1)}_{\nu}(k\tau)\,H^{(2)}_{\nu}(k\tau). \label{ReA2}\eeq
The quantum state specified in this manner is known as the Bunch-Davies
vacuum \cite{BuDa,Allen}.\\
Using (\ref{Bew4}), (\ref{Bew5}) and (\ref{ReA2}) it turns out that the
integrals (\ref{EIT4}) involving two Bessel functions are of the
Weber-Schafheitlin type and can be evaluated analytically. The results are
given in appendix \ref{Hankel}. Expansion in $\e=3\!-\!d$ yields
($\psi(x)=\G'(x)/\G(x)$)
\beq \<\Psi|T_{\mu\nu}|\Psi\>\stackrel{\e\to0}{=}
 {-g_{\mu\nu}\0d\!+\!1}\,{m^2H_0^2\016\pi^2}\Bigl(\4-\nu^2\Bigr)
 \biggl({2\0\e}-\g+1+\ln{4\pi\0H_0^2}-{\ts\psi({3\02}\!+\!\nu)
 -\psi({3\02}\!-\!\nu)}+{\cal O}(\e)\biggr). \label{SpT2}\eeq
The $1/\e\,$-pole can be removed by a renormalization through the subtraction
of De~Witt-Schwinger terms as described in the foregoing section, and we
obtain the final result:
\bea \<T_{\mu\nu}\>_{\rm ren}
 &=&{1\064\pi^2}\,g_{\mu\nu}\Bigl(m^2(m^2-{\ts(\xi\!-\!{1\06})}R)
 \Bigl({\ts\psi({3\02}\!+\!\nu)+\psi({3\02}\!-\!\nu)}+\ln{H_0^2\0m^2}\Bigr) \nn\\
 & &\qqqd+\;m^2{\ts(\xi\!-\!{1\06})}R+{\ts{1\018}}\,m^2R-a_2\Bigr)
 \label{EIT5}\eea
with $a_2=-{1\02160}\,R^2+\2(\xi\!-\!{1\06})^2R^2\,$, $R=-12\,H_0^2\,$.\\
Due to the de~Sitter symmetry this expectation value is proportional to the
metric tensor and acts just like an effective cosmological constant within
the Einstein equations. Therefore a de~Sitter spacetime with its Hubble
constant $H_0$ determined by the transzendental equation
\beq 3\,H_0^2+\La\;=\;8\pi G_N\,\r_{\rm ren}(H_0^2) \label{selk}\eeq
together with the Bunch-Davies vacuum forms a solution of the semiclassical,
\hbox{coupled} system of equations (\ref{EinSchro}).\\
The solutions of (\ref{selk}) for given $\La$ have been studied in ref.
\cite{WadAz}. Clearly, for every given $H_0$ there is a $\La$ so that (\ref{selk})
is fulfilled. Thus a de~Sitter spacetime of arbitrary curvature is possible.
\expandafter\ifx\csname ok\endcsname\relax
   \end{document}\fi
\expandafter\ifx\csname qd\endcsname\relax
   \documentstyle[12pt,twoside]{article}\let\qd=\quad \let\qqd=\qquad \def\qqqd{\qquad\qquad}

\let\a=\alpha \let\be=\beta \let\g=\gamma \let\de=\delta
\let\e=\varepsilon \let\z=\zeta \let\h=\eta \let\th=\theta
\let\dh=\vartheta \let\k=\kappa \let\la=\lambda
\let\r=\rho \let\s=\sigma
\let\om=\omega
\let\ph=\varphi
\let\Om=\Omega \let\Si=\Sigma \let\Th=\Theta
\let\La=\Lambda \let\G=\Gamma \let\D=\Delta

\def\0{\over } \def\1{\vec } \def\2{{1\over2}} \def\4{{1\over4}}
\def\5{\bar } \def\6{\partial } \def\7#1{{#1}\llap{/}}
\def\8{\tilde } \def\9{\dot }

\def\ul#1{$\underline{\hbox{#1}}$} \def\llp{\hbox to 0pt{\hss /\hskip1.5pt}}
\def\so{\supset\hbox to 0pt{\hss $\displaystyle -$\hskip1pt}}

\def\ts{\textstyle } \def\ds{\displaystyle }
\def\<{\langle } \def\>{\rangle } \def\lb{\left} \def\rb{\right}
\let\lra=\leftrightarrow \let\LRA=\Leftrightarrow
\let\then=\Rightarrow \def\triang{{\scriptstyle \Delta}}
\def\trian{{\scriptscriptstyle \Delta}}
\def\re{{\rm Re}} \def\im{{\rm Im}} \def\ord{{\cal O}}

\let\ap=\approx \let\eq=\equiv \let\ex=\times \let\hc=\dagger
      \let\ti=\tilde \let\bl=\biggl \let\br=\biggr
\let\bi=\choose \let\at=\atop \let\mat=\pmatrix

\let\nn=\nonumber
\def\bea{\begin{eqnarray}} \def\eea{\end{eqnarray}}
\def\bean{\begin{eqnarray*}} \def\eean{\end{eqnarray*}}
\def\beq{\begin{equation}} \def\eeq{\end{equation}}
\def\barr{\begin{array}} \def\earr{\end{array}}

\newcommand{\deflabel}[1]{#1\hfil}%
\newenvironment{deflist}[1]%
{\begin{list}{}%
{\settowidth{\labelwidth}{#1}%
\setlength{\leftmargin}{\labelwidth}\addtolength{\leftmargin}{\labelsep}%
\renewcommand{\makelabel}{\deflabel}}}%
{\end{list}}%
\fi
\expandafter\ifx\csname ok\endcsname\relax
   \def\re{{\rm Re}} \def\im{{\rm Im}}
   {\catcode`@=11 \@addtoreset{equation}{section}\@addtoreset{figure}{section}}
   \def\theequation{\thesection.\arabic{equation}}
   \def\thefigure{\thesection.\arabic{figure}}
   \addtolength{\topmargin}{-48pt}
   \addtolength{\textheight}{90pt}
   \addtolength{\evensidemargin}{-31pt}
   \addtolength{\oddsidemargin}{-12pt}
   \addtolength{\textwidth}{30pt}
   \addtolength{\footskip}{21pt}
   \setlength{\parindent}{0pt}\frenchspacing
   \includeonly{}
   \begin{document}
   
\fi
\section{Nearly de~Sitter spacetimes}\label{Fast}
We are now approaching our main goal: the linear stability analysis of the
semiclassical solution from the foregoing chapter. Our starting point is
that of reference \cite{TraHill}: We will consider small fluctuations of
the gravitational field and small perturbations of the matter quantum state.
The semiclassical equations are linearized around the
de~Sitter -- Bunch-Davies solution. At time $t_0$ the whole system will be
given an initial configuration, which slightly deviates from de~Sitter
spacetime and Bunch-Davies vacuum. Then we analyse the time evolution of
this deviation. Any instabilities would be indicated by (exponentially)
growing parts in the general solution for the deviation.\\
For simplicity and feasibility only fluctuations of the metric will be
considered, which do not break its spatial homogeneity and isotropy.
This is of course a limitation, but it has already been shown (for example
in reference \cite{HaHu}) that small initial anisotropies are damped away
by particle production and an automatic isotropization takes place.\\
The Gaussian form of the wave functional is not altered by the metric
fluctuations. This is also assumed for its initial
deviation. Due to the linearization this assumption does not exclude an
initial wave functional containing first excitations. Moreover the energy
density of initial excitations would be subject to the
exponential de~Sitter red-shift and have no influence on the long term
behaviour of the system.\\
Firstly we want to compute the change in the Gaussian width of the wave
functional and in the energy momentum components (\ref{EIT4}) for a given
fluctuation of the FRW scale factor and initial deviation from the
Bunch-Davies vacuum. In the sequel quantities related to the unperturbed
de~Sitter spacetime and Bunch-Davies vacuum will get the index 0,
whereas a prefix $\de$ always means the deviation of a quantity
from its unperturbed value.\\ In order to save some ink the Hubble parameter
$H_0$ of the unperturbed de~Sitter spacetime will be set equal to 1 (in
addition to $\hbar$ and $c$). This means that masses are measured in units of
$H_0$.\\
Since the results do not depend on the starting time, $t_0=0$ will be used
without loss of generality.\\
Consider now a small deviation of the FRW scale factor from its de~Sitter
value $a_0(t)=e^t\,$:
\bea a(t)&=&a_0(t)\,(1+I(t))\,,\qqd I(t)\ll1 \nn\\
    H(t)&=&{\9a\0a}\;=\;1+\9I(t) \label{Skal}\eea
In the following every quantity will be linearized with respect to $I(t)$.
The components of the energy momentum tensor are written as
\bean \r(t)&=&\<T_{00}\>\;=\;\r_0+\de\r(t) \\
  p(t)&=&-{1\0d}\,g^{ij}\<T_{ij}\>\;=\;p_0+\de p(t)\;,\eean
where $\r_0$ and $p_0$ are the unperturbed quantities (\ref{EIT5}). There are two
sources of contributions to $\de\r$ und $\de p$: The first one emerges from the
explicit appearance of the metric in the definition of $T_{\mu\nu}\,$, and the
other one is due to the dependence of the Hamiltonian on the metric leading
to a deviation of the wave functional from the Bunch-Davies vacuum:
\[ A(k,t)\;=\;A_0(k,t)+\de A(k,t) \]
With the help of (\ref{Bew4}), (\ref{Bew5}) and by noting that
$\6_t\int\8d\8k/(2\,\re A_0(k,t))=0$ we obtain from (\ref{EIT4}):
\bea \de\r&=&\xi\,(\de R_{00}\!-\!\de R)\int\!\8d\8k\,{1\02\,\re A_0}\;
 +\int\!\8d\8k\,k^2\de\biggl({a^{-2}\02\,\re A}\biggr)\nn\\ &&+\;\Bigl(m^2+\xi\,
  (R_{00}\!-\!R)+d\,({\ts\xi\!+\!\4})\,\6_t+{\ts\4}\,\6_t^2\Bigr)
  \int\!\8d\8k\,\de\biggl({1\02\,\re A}\biggr)\nn\\
  \de p&=&\xi\,\de(a^{-2}R_{11})\int\!\8d\8k\,{1\02\,\re A_0}\;
  +{1\0d}\int\!\8d\8k\,k^2\de\biggl({a^{-2}\02\,\re A}\biggr)\nn\\
  & &+\;\Bigl(\xi\,a^{-2}R_{11}
 +({\ts{d\04}+\xi(1\!-\!d)})\,\6_t+({\ts\4\!-\!\xi})\,\6_t^2\Bigr)\int\!\8d\8k\,
  \de\biggl({1\02\,\re A}\biggr) \label{EIT6}\eea
Since $\re A(k,t)=\sqrt{g(t)}\,\re\,\G(k,t)$ we have
\beq \de\biggl({1\02\,\re A}\biggr)\;=\;-{d\,I\02\,\re A_0}-F_k\;,\qd
  \de\biggl({a^{-2}\02\,\re A}\biggr)\;=\;-a_0^{-2}{(d\!+\!2)\,I\02\,\re A_0}
  -a_0^{-2}F_k \label{Fk}\eeq
with $\ds F_k:=-a_0^{-d}\de\biggl({1\02\,\re\,\G}\biggr)\,$.\\
The Schr\"odinger equation (\ref{Bew2}) leads to the following equation of motion
for $\de\G(k,t)$:
\beq i\,\6_t\de\G(k,t)-2\,\G_0(k,t)\,\de\G(k,t)\;=\;2\,I(t)\,a_0^{-2}k^2+r(t)
  \label{Bew7}\eeq
with
\bea r(t)&:=&{d^2\02}\,\9I(t)+{d\02}\,\ddot I(t)+\xi\,\de R(t)\;=\;
  \2\,\9I(t)+\2\,\ddot I(t)+\xi_c\de R(t)\;,\nn\\
  \xi_c&:=&\xi-{d-1\04d}\;. \label{r0}\eea
From the foregoing section we know that
$\G_0(k,t)=-i\,\6_t\ln H^{(2)}_{\nu}(k\tau_0(t))$, $\tau_0(t)=a_0^{-1}=e^{-t}$,
so that the general solution of (\ref{Bew7}) is
\bea \de\G(k,t)&=&-\;{i\0{H^{(2)}_{\nu}}^2\!(k\tau_0(t))}\,\int\limits_0^t\!dt'\,
\Bigl(r(t')+2\,I(t')\,a_0^{-2}\!(t')\,k^2\Bigr)\,{H^{(2)}_{\nu}}^2\!(k\tau_0(t'))\nn\\
  &&+\;{{H^{(2)}_{\nu}}^2\!(k)\0{H^{(2)}_{\nu}}^2\!(k\tau_0(t))}\;
  \de\G(k,0) \eea
Using this and (\ref{ReA2}) we find
\bea F^{\phantom{(ii)}}_k\!\!&=&\!a_0^{-d}{\re\,\de\G\02\,(\re\,\G_0)^2}\;=\;
  F^{(i)}_k+F^{(ii)}_k \nn\\
  F^{(i)\phantom{i}}_k\!\!&:=&\!a_0^{-d}\,\re\biggl(-i\,{\pi^2\08}\,
  {H^{(1)}_{\nu}}^2\!(k\tau_0(t))\int\limits_0^t\!dt'\,\Bigl(r(t')+
2\,I(t')\,a_0^{-2}\!(t')\,k^2\Bigr)\,{H^{(2)}_{\nu}}^2\!(k\tau_0(t'))\biggr)\nn\\
  F^{(ii)}_k\!\!&:=&\!a_0^{-d}\,\re\biggl({\pi^2\08}\,{H^{(1)}_{\nu}}^2\!(k\tau_0(t))\,
  {H^{(2)}_{\nu}}^2\!(k)\,\de\G(k,0)\biggr)\;. \label{Fk1}\eea
The initial deviation $\de\G(k,0)$ from the Bunch-Davies vacuum $\G_0(k,0)$ is
part of the initial data of the problem. Since a physically meaningful, perturbed
initial state should have a finite energy density, $\de\G(k,0)$ must have a special
high energy behaviour:
\beq \de\G(k,0)\;=\;\de\G^{(ii)}(k)-i\sum_{n=-3}^{+1}(ik)^n\de\G_n\;,
  \label{dGam}\eeq
where $\de\G^{(ii)}(k)\stackrel{k\to\infty}{\to}0$ faster than $k^{-3}$, and
the coefficients $\de\G_n$ are determined by a comparison with the
adiabatic vacuum in the next section. Of course $\de\G(k,0)$ has to be finite
for $k\to 0$. This can be ensured by suitable $\de\G^{(ii)}(k)$ in the form (\ref{dGam}).

\subsection{The adiabatic vacuum}\label{adia}
Although it is not possible to define a unique vacuum state in a general
curved background spacetime, it is possible to define a
state which is vacuos for the high $k\,$-modes in the limit $k\to\infty$.
This can be achieved by using the adiabatic expansion of positive frequency
for the field modes, which is at the same time an expansion in $k^{-1}$
and becomes exact in the above limit. We require the width $\G(k,0)$ of our
quantum state to coincide with the width $\G_{\rm ad}(k,0)$ of the adiabatic
vacuum in the limit of large $k$, so our energy momentum tensor expectation
value will have the same divergencies as the De~Witt-Schwinger one (being
also a local expansion). This yields a finite renormalized energy momentum
density.\\
Since the energy momentum expectation value is quartically divergent, the
terms up to a relative order of $k^{-4}$ in the Gaussian width
$\G_{\rm ad}(k,t)\sim k$ are responsible for its divergencies. ``In the limit
of large $k$'' means therefore ``up to the relative order of $k^{-4}$ in the
limit $k\to\infty$''.\\
We are now proceeding with the computation of $\G_{\rm ad}(k,t)$ by an
adiabatic expansion (positive frequency) of the solution of (\ref{Bew3}).
Again the conformal time
$\tau:=-\int_{t_0}^tdt'/a(t')$ with $dt^2=a^2(t)\,d\tau^2$ is introduced.
We substitute $u_{\rm ad}(k,t)=:a^{1/2}(t)\,\chi_k(\tau)$ and
(\ref{Bew3}) takes the form
\beq \6_{\tau}^2\,\chi_k(\tau)+\Om_k^2(\tau)\,\chi_k(\tau)\;=\;0 \label{ad1}\eeq
with $\Om_k(\tau):=(k^2+a^2\8m^2)^{1/2}\,,\qqd\8m^2:=m^2-\xi_cR\,.$\\
The adiabatic solutions of positive frequency are
\beq \chi_k(\tau)\;=\;{1\0\sqrt{2\,W_k(\tau)}}\,
\exp\Bigl(-i\int\limits_{\5{\tau}}^{\tau}\!d\tau'\,W_k(\tau')\Bigr)\qd,\label{ad2}\eeq
where the $W_k$ have to obey the following equations:
\beq W_k^2+\2\,{W_k''\0W_k}-{3\04}\,{W_k'^2\0W_k^2}\;=\;\Om_k^2 \eeq
These are solved iteratively order by order ($'=d/d\tau$):
\bean W_k^{\phantom{(0)}}&=&W_k^{(0)}+W_k^{(2)}+W_k^{(4)}
  +\;\cdots\phantom{{\Om_k'^2\0\Om_k^2}}\\
  W_k^{(0)}&=&\Om_k\phantom{{\Om_k'^2\0\Om_k^2}}\\
  W_k^{(2)}&=&{3\08}\,{\Om_k'^2\0\Om_k^3}-\4\,{\Om_k''\0\Om_k^2}\\
  W_k^{(4)}&=&{1\016}\,{\Om_k^{\rm(iv)}\0\Om_k^4}-{5\08}\,{\Om_k'''\Om_k'\0
  \Om_k^5}-{13\032}\,{\Om_k''^2\0\Om_k^5}+{99\032}\,{\Om_k''\Om_k'^2\0\Om_k^6}
  -{297\0128}\,{\Om_k'^4\0\Om_k^7} \eean
Using the above $\Om_k$ and expanding with respect to $k^{-1}$ up to the fourth
order relative to the leading one we obtain:
\beq W_k\;=\;k+{a^2\0k}\,{\8m^2\02}-{a^4\0k^3}\,{1\08}\,(\8m^4+
 2\8m^2(\9H\!+\!3H^2)-\xi_c(5H\9R\!+\!\ddot R))+{\cal O}(k^{-4})\label{ad4}\eeq
$W_k^{(4)}$ and the first term in $W_k^{(2)}$ are already of the relative order
of $k^{-5}$ and do not appear in (\ref{ad4}). Using $dt/d\tau=-a(t)$ the
derivatives with respect to $\tau$ have been converted to those with respect
to $t$. Putting things together we get the adiabatic width:
\bea \hbox to 0pt{\hss$\G_{\rm ad}$}(k,t)\!&=&\!
  -i\,{\9u_{\rm ad}(k,t)\0u_{\rm ad}(k,t)}\;=\;
  {1\0a}\,W_k-i\,\Bigl({H\02}-\2\,{\9W_k\0W_k}\Bigr) \nn\\
  \!&=&\!a^{-1}k-i\,{H\02}+{1\0a^{-1}k}\,{\8m^2\02}+{1\0a^{-2}k^2}\,{i\04}\,
  (2\,H\,\8m^2-\xi_c\9R)\nn\\
  &&\!-\;{1\0a^{-3}k^3}\,{1\08}\,(\8m^4+2\8m^2(\9H\!+\!3H^2)-\xi_c(5H\9R\!+\!
  \ddot R))+{\cal O}(k^{-4})\phantom{M}\label{ad5}\eea
This has to be compared with our Bunch-Davies width $\G_0(k,t)$.
\[ \G_0(k,t)\;=\;-i\,{\9u_0(k,t)\0u_0(k,t)}\;=\;
  -i\,{\6_tH^{(2)}_{\nu_o}(k\tau_0)\0H^{(2)}_{\nu_o}(k\tau_0)}\;=\;
 ia_0^{-1}k\,{\6_{k\tau_o}H^{(2)}_{\nu_o}(k\tau_0)\0H^{(2)}_{\nu_o}(k\tau_0)}\]
Using the asymptotic expansion of the Hankel functions
\[ {{H^{(2)}_{\nu}}'(z)\0H^{(2)}_{\nu}(z)}\;\stackrel{|z|\to\infty}{=}\;
 -i-{1\02z}-i\,{\4\!-\!\nu^2\02z^2}+{\4\!-\!\nu^2\02z^3}+i\,{(\4\!-\!\nu^2)
 ({25\04}\!-\!\nu^2)\08z^4}+{\cal O}(z^{-5}) \]
as well as $\4-\nu_0^2=\8m_0^2/H_0^2$ we obtain:
\beq \G_0(k,t)=a^{-1}k-{i\02}H_0+{1\0a_0^{-1}k}{\8m_0^2\02}+
  {i\0a_0^{-2}k^2}{H_0\8m_0^2\02}-{1\0a_0^{-3}k^3}{\8m_0^2\08}(\8m_0^2
  +6H_0^2)+{\cal O}(k^{-4})\label{ad0}\eeq
For the sake of clarity we did not replace $H_0$ by 1 in this formula.
The asymptotic expansion (\ref{ad0}) coincides with (\ref{ad5}) up to the order
given in the special case of de~Sitter spacetime ($\9H_0=\9R_0=\ddot R_0=0$):
$\G_0=\G_{\rm ad\,0}+{\cal O}(k^{-4})$. This means that $B_1(k\to\infty)\to 0$
was the correct choice in section \ref{DeSit}.\\
For the nearly de~Sitter spacetime we have to require
$\de\G(k,0)+\G_0(k,0)=\G_{\rm ad}(k,0)+{\cal O}(k^{-4})$. After linearizing
(\ref{ad5}) with respect to the deviation from the de~Sitter spacetime we are
finally in the position to obtain the coefficients $\de\G_n$ needed in
(\ref{dGam}):
\bea \de\G_1&=&-I(0)\,,\qd\de\G_0\;=\;\2\,\9I(0)\,,\qd\de\G_{-1}\;=\;
  -\2\,(\8m_0^2I(0)-\xi_c\de R(0))\,,\nn\\
\de\G_{-2}&=&\4\,(4\8m_0^2I(0)+2\8m_0^2\9I(0)-\xi_c(2\de R(0)\!+\!\de\9R(0)))\,,
\nn\\
  \de\G_{-3}&=&-{1\08}\,(3I(0)(\8m_0^4\!+\!6\8m_0^2)+2\8m_0^2(\ddot I(0)\!+\!
  6\9I(0)\!-\!\xi_c\de R(0))\nn\\
  &&\phantom{-{1\08}\,(}-\xi_c(6\de R(0)\!+\!5\de\9R(0)\!+\!\de\ddot R(0)))
  \,.\label{dGamn}\eea

\expandafter\ifx\csname ok\endcsname\relax
   \end{document}\fi

\expandafter\ifx\csname qd\endcsname\relax
   \documentstyle[12pt,twoside]{article}\let\qd=\quad \let\qqd=\qquad \def\qqqd{\qquad\qquad}

\let\a=\alpha \let\be=\beta \let\g=\gamma \let\de=\delta
\let\e=\varepsilon \let\z=\zeta \let\h=\eta \let\th=\theta
\let\dh=\vartheta \let\k=\kappa \let\la=\lambda
\let\r=\rho \let\s=\sigma
\let\om=\omega
\let\ph=\varphi
\let\Om=\Omega \let\Si=\Sigma \let\Th=\Theta
\let\La=\Lambda \let\G=\Gamma \let\D=\Delta

\def\0{\over } \def\1{\vec } \def\2{{1\over2}} \def\4{{1\over4}}
\def\5{\bar } \def\6{\partial } \def\7#1{{#1}\llap{/}}
\def\8{\tilde } \def\9{\dot }

\def\ul#1{$\underline{\hbox{#1}}$} \def\llp{\hbox to 0pt{\hss /\hskip1.5pt}}
\def\so{\supset\hbox to 0pt{\hss $\displaystyle -$\hskip1pt}}

\def\ts{\textstyle } \def\ds{\displaystyle }
\def\<{\langle } \def\>{\rangle } \def\lb{\left} \def\rb{\right}
\let\lra=\leftrightarrow \let\LRA=\Leftrightarrow
\let\then=\Rightarrow \def\triang{{\scriptstyle \Delta}}
\def\trian{{\scriptscriptstyle \Delta}}
\def\re{{\rm Re}} \def\im{{\rm Im}} \def\ord{{\cal O}}

\let\ap=\approx \let\eq=\equiv \let\ex=\times \let\hc=\dagger
      \let\ti=\tilde \let\bl=\biggl \let\br=\biggr
\let\bi=\choose \let\at=\atop \let\mat=\pmatrix

\let\nn=\nonumber
\def\bea{\begin{eqnarray}} \def\eea{\end{eqnarray}}
\def\bean{\begin{eqnarray*}} \def\eean{\end{eqnarray*}}
\def\beq{\begin{equation}} \def\eeq{\end{equation}}
\def\barr{\begin{array}} \def\earr{\end{array}}

\newcommand{\deflabel}[1]{#1\hfil}%
\newenvironment{deflist}[1]%
{\begin{list}{}%
{\settowidth{\labelwidth}{#1}%
\setlength{\leftmargin}{\labelwidth}\addtolength{\leftmargin}{\labelsep}%
\renewcommand{\makelabel}{\deflabel}}}%
{\end{list}}%
\fi
\expandafter\ifx\csname ok\endcsname\relax
   \def\re{{\rm Re}} \def\im{{\rm Im}}
   {\catcode`@=11 \@addtoreset{equation}{section}\@addtoreset{figure}{section}}
   \def\theequation{\thesection.\arabic{equation}}
   \def\thefigure{\thesection.\arabic{figure}}
   \addtolength{\topmargin}{-48pt}
   \addtolength{\textheight}{90pt}
   \addtolength{\evensidemargin}{-31pt}
   \addtolength{\oddsidemargin}{-12pt}
   \addtolength{\textwidth}{30pt}
   \addtolength{\footskip}{21pt}
   \setlength{\parindent}{0pt}\frenchspacing
   \includeonly{}
   \begin{document}
   
\fi

\subsection{Momentum integrals and isolation of divergencies}\label{Div}
\def\vFd{{}_4\5F_3}
\def\vFe{{}_4\8F_3}
\def\dI{I\raise10.1pt\hbox to 0pt{\hss.\hskip-1.44pt.\hskip-1.44pt.\hskip-1.8pt}}
\def\dr{r\raise7.0pt\hbox to 0pt{\hss.\hskip-1.44pt.\hskip-1.44pt.\hskip-0.8pt}}
If we insert (\ref{Fk}) into equation (\ref{EIT6}) there are two integrals of the
Weber-Schafheitlin type involving two Hankel functions, which already appeared
in section \ref{DeSit}. After substituting $y:=k\tau$ they are evaluated using
(\ref{WebSchaf3}):
\bea J^{(2)}_0&:=&\int\!\8d\8k\,{1\02\,\re A_0(k,t)}\nn\\
  &=&-{\8m_0^2\016\pi^2}\biggl({2\0\e}-\g+1+\ln4\pi-{\ts\psi({3\02}\!+\!\nu_0)
  -\psi({3\02}\!-\!\nu_0)}+{\cal O}(\e)\biggr)\nn\\
  J^{(2)}_2&:=&\int\!\8d\8k\,{a_0^{-2}(t)\,k^2\02\,\re A_0(k,t)}\label{Int0}\\
  &=&{\8m_0^2\016\pi^2}\,{3\04}\,(m^2\!-\!\xi R_0)\biggl({2\0\e}-\g+{5\06}
 +\ln4\pi-{\ts\psi({3\02}\!+\!\nu_0)-\psi({3\02}\!-\!\nu_0)}+{\cal O}(\e)\biggr)
  \nn\eea
The other integrals appearing in (\ref{EIT6}) are involving four Hankel
functions. In terms of\\
\parbox{\textwidth}{
\bea J^{(4)}_l(t\!-\!t')&:=&a_0^{-d}(t)\int\!\8d\8k\,(k\tau_0(t'))^l\,
 {H^{(1)}_{\nu}}^2\!(k\tau_0(t))\,{H^{(2)}_{\nu}}^2\!(k\tau_0(t'))\nn\\
 &=&{2^{1-d}\tau_0^d(t\!-\!t')\0\G({d\02})\,\pi^{d/2}}\int\limits_0^{\infty}\!dy\,
 y^{l+d-1}\,{H^{(1)}_{\nu}}^2\!(y\tau_0(t\!-\!t'))\,{H^{(2)}_{\nu}}^2\!(y)
 \label{J4l}\eea
}
they explicitly read
\bea \int\!\8d\8k\,F^{(i)}_k&=&\re\biggl(-i\,{\pi^2\08}\int\limits_0^t\!dt'\,
 \Bigl(r(t')\,J^{(4)}_0(t\!-\!t')+2I(t')\,J^{(4)}_2(t\!-\!t')\Bigr)\biggr)\label{Fki}\\
 \int\!\8d\8k\,a_0^{-2}k^2F^{(i)}_k&=&\re\biggl(-i\,{\pi^2\08}\int\limits_0^t\!dt'\,
 \tau_0^2(t\!-\!t')\,\Bigl(r(t')\,J^{(4)}_2(t\!-\!t')+2I(t')\,J^{(4)}_4(t\!-\!t')
 \Bigr)\biggr)\nn\\
 \int\!\8d\8k\,F^{(ii)}_k&=&\re\biggl(-i\,{\pi^2\08}\sum_{n=-3}^1\de\G_n\,i^n
 J^{(4)}_n(t)\biggr)+\de\G^{(ii)}\,\hbox{-terms}\label{Fkii}\\
 \int\!\8d\8k\,a_0^{-2}k^2F^{(ii)}_k&=&\re\biggl(-i\,{\pi^2\08}\,\tau_0^2(t)
 \sum_{n=-3}^1\de\G_n\,i^nJ^{(4)}_{n+2}(t)\biggr)+\de\G^{(ii)}\,\hbox{-terms}\,.
 \nn\eea
The $t'\,$-integrations in (\ref{Fki}) are convolution integrals. The fact that
the $J^{(4)}_l$ defined above depend only on the difference $t\!-\!t'$ means
independence of the starting time ($t_0=0$ here) and follows from the maximum
symmetry of de~Sitter spacetime.\\
The integrals (\ref{J4l}) involving a product of four Hankel resp. Bessel
functions were not found in the mathematical standard literature. Therefore
their evaluation has been included as part of this work in appendix
\ref{Hankel}.\\
At this point the investigations in references
\cite{TraHill} and \cite{GuvLieb} failed. In order to circumvent the
integrations of four Hankel functions they carry out a so called
``short-time'' approximation, which consists in a restriction on short times
$t$ and the Taylor series expansion of $J^{(4)}_l(t\!-\!t')$ around $t'=t$ up to
the linear order. However, the $J^{(4)}_l(t\!-\!t')$ have a singularity at
$t'=t$ (see below) and cannot be expanded around this point. Hence this
approximation does not lead to correct results even for arbitrarily short
times. Moreover the divergencies are not coming out correctly, so no
sensible renormalization is possible. Comparing with our exact results
(\ref{ex1}) and (\ref{ex2}) it turns out that even the leading terms of a real
short-time approximation are missing.\\
In reference \cite{IsaRog} the momentum
integrations are not executed and one is left with even more complicated
integrals in the final result.\\
We show in appendix \ref{Hankel} Eq. (\ref{AJ4l}) that
\beq J^{(4)}_l(t)\;=\;-{2^{3-d}(-i)^{d+l}\0\G({d\02})\,\pi^{2+d/2}}\,\,G(t;0,l)
 \;.\label{J4l1}\eeq
The dependence of $G$ and $J^{(4)}_l$ on $d$ and $\nu$ has been suppressed in
favour of a shorter notation. The definition of $G$ in appendix \ref{Hankel} explictly reads:
\bea \lefteqn{G(t;p,l)\;:=\;{-\tau_0^d(t)\04\sin^2\pi\nu}\biggl(}\nn\\
  &&\!\!-\;2\;\vFd\Bigl(\hbox{\small$\ds
  {d\!+\!l\02}\!-\!p,{d\!+\!l\02}\!-\!\nu,{d\!+\!l\02}\!+\!\nu,\2;
  1\!+\!\nu,1\!-\!\nu,{d\!+\!l\!+\!1\02};\tau_0^2(t)$}\Bigr)\nn\\
  &&\!\!+\,\tau_0^{-2\nu}\!(t)\,e^{2\pi i\nu}\vFd
  \Bigl(\hbox{\small$\ds{d\!+\!l\02}\!-\!\nu\!-\!p,
  {d\!+\!l\02}\!-\!2\nu,{d\!+\!l\02},\2\!-\!\nu;1\!-\!\nu,1\!-\!2\nu,
  {d\!+\!l\!+\!1\02}\!-\!\nu;\tau_0^2(t)$}\Bigr)\nn\\
  &&\!\!+\,\tau_0^{2\nu}\!(t)\,e^{-2\pi i\nu}\vFd
  \Bigl(\hbox{\small$\ds{d\!+\!l\02}\!+\!\nu\!-\!p,
  {d\!+\!l\02},{d\!+\!l\02}\!+\!2\nu,\2\!+\!\nu;1\!+\!\nu,1\!+\!2\nu,
  {d\!+\!l\!+\!1\02}\!+\!\nu;\tau_0^2(t)$}\Bigr)\biggr)\nn\\
  \label{G1}\eea
Our generalized hypergeometric function $\vFd$ appearing in (\ref{G1}) is defined
by an infinite series:
\bea \lefteqn{\vFd(\a_1,\a_2,\a_3,\a_4;\be_1,\be_2,\be_3;z)\;:=}\hskip60pt\nn\\
 &&{\G(\a_1)\,\G(\a_2)\,\G(\a_3)\,\G(\a_4)\0\G(\be_1)\,\G(\be_2)\,\G(\be_3)}\;
 {}_4F_3(\a_1,\a_2,\a_3,\a_4;\be_1,\be_2,\be_3;z)\nn\\
 &=&\sum_{n=0}^{\infty}{\G(\a_1\!+\!n)\,\G(\a_2\!+\!n)\,\G(\a_3\!+\!n)\,
 \G(\a_4\!+\!n)\0n!\;\G(\be_1\!+\!n)\,\G(\be_2\!+\!n)\,\G(\be_3\!+\!n)}\;z^n\;,
 \label{F1}\eea
where ${}_pF_q$ is the function usually called generalized hypergeometric
function in the mathematical literature.\\
According to appendix \ref{Hankel} the integral (\ref{J4l}) is convergent for
$\tau_0(t\!-\!t')\neq1$ resp. $t'\neq t$ in the region $4\,|\re\,\nu|<l+\re\,d<3$
(which has always a non-zero extension if $m^2\!-\!\xi R_0>0$). Equation
(\ref{J4l1}) gives its analytic continuation on the whole
complex $d\,$-plane.\\
In order to investigate the convergence behaviour of the series (\ref{F1}),
we need an asymptotic expansion of its terms for large $n$. This can be
obtained using Stirling's series for the gamma-function
(see \cite[\S13.6]{WhittWat}):
\beq \G(\a+n)\;\stackrel{n\to\infty}{=}\;\sqrt{2\pi}\,e^{-n}n^{n+\a-\2}
 \exp\biggl(\sum_{m=1}^M{(-)^{m+1}B_{m+1}(\a)\0m(m+1)\,n^m}+
 {\cal O}\Bigl({1\0n^{M+1/2}}\Bigr)\biggr), \label{Stirl}\eeq
where $B_m(\a)$ are the Bernoulli polynoms:
\bea &&B_0(x)=1,\ B_1(x)=x-\2\,,\ B_2(x)=x^2-x+{1\06}\,,\ B_3(x)=x^3-{3\02}x^2
 +{x\02}\,,\nn\\
 &&B_4(x)=x^4-2x^3+x^2-{1\030}\,,\ B_5(x)=x^5-{5\02}x^4+{5\03}x^3-{1\06}x\,,\
 \ldots\label{Bern}\eea
Defining
\bea \5B_m(\a_1,\a_2,\a_3,\a_4;\be_1,\be_2,\be_3,\be_4)&:=&
 \sum_{i=1}^4B_m(\a_i)-\sum_{i=1}^4B_m(\be_i)\,,\nn\\
 \s&:=&\sum_{i=1}^4\a_i-\sum_{i=1}^3\be_i-1 \label{Bbar}\eea
we get
\bea Q_n&:=&{\G(\a_1\!+\!n)\,\G(\a_2\!+\!n)\,\G(\a_3\!+\!n)\,\G(\a_4\!+\!n)\0
 \G(\be_1\!+\!n)\,\G(\be_2\!+\!n)\,\G(\be_3\!+\!n)\,\G(1\!+\!n)}\nn\\
 &\stackrel{n\to\infty}{=}&n^{\s}
 \exp\biggl(\sum_{m=1}^M{(-)^{m+1}\5B_{m+1}(\a_1,\a_2,\a_3,\a_4;\be_1,\be_2,
 \be_3,1)\0m(m+1)\,n^m}+{\cal O}\Bigl({1\0n^{M+1/2}}\Bigr)\biggr)\nn\\
 &\stackrel{M=4}{=}&n^{\s}\biggl(1+{\8B_1(\ldots)\0n}+
 {\8B_2(\ldots)\0n^2}+{\8B_3(\ldots)\0n^3}
 +{\8B_4(\ldots)\0n^4}+{\cal O}\Bigl({1\0n^{9/2}}\Bigr)\biggr)\,.\label{Qn}\eea
The polynoms $\8B_m$ arise from an expansion of the exponential function and
are polynoms of Bernoulli polynoms:
\bea &&\8B_1(\ldots):={\5B_2(\ldots)\02}\,,\ \8B_2:={\5B_2^2\08}-{\5B_3\06}\,,\
 \8B_3:={\5B_2^3\048}-{\5B_2\5B_3\012}+{\5B_4\012}\nn\\
 &&\8B_4:={\5B_2^4\0384}-{\5B_2^2\5B_3\048}+{\5B_2\5B_4\024}+{\5B_3^2\072}
 -{\5B_5\020}\,,\ \8B_0:=1\;. \label{Btil}\eea
The series (\ref{F1}) is always convergent for $0\leq z<1$ respectively $t>0$ in
(\ref{G1}).\\
Putting in the arguments $\a_1\ldots\be_3$ from (\ref{G1}) we find $\s=d+l-3-p$
and therefore $0\leq\s\leq4$ for $d=3$, $l=0,2,4$ and $p=0$. This means that
in the limit $z\to1$ ($t\to0$ in (\ref{G1})) the leading terms of the series are
the ones of large $n$, which can be computed using the expansion (\ref{Qn}).\\
Introducing the function
\beq F(z,s)\;=\;\sum_{n=1}^{\infty}{z^n\0n^s}\qd\eeq
we obtain
\bea \lefteqn{\vFd(\a_1,\ldots,\a_4;\be_1,\ldots,\be_3;z)\;=\;\sum_nQ_nz^n}
 \hskip60pt\nn\\
 &\stackrel{z\to1}{=}&F(z,-\s)+\8B_1\,F(z,-\s\!+\!1)+\8B_2\,F(z,-\s\!+\!2)\nn\\
 &&+\;\8B_3\,F(z,-\s\!+\!3)+\8B_4\,F(z,-\s\!+\!4)+\,\ldots\;. \label{Fas}\eea
According to \cite[chapter 1.11]{HTI} the function $F$ obeys the following relations
($B_m=B_m(0)$ Bernoulli numbers):
\bea F(z,-m)&=&m!\,(-\ln z)^{-m-1}-\sum_{r=0}^{\infty}
 {B_{m+r+1}\0(m+r+1)\,r!}\,(\ln z)^r\,,\qqd m=1,2,3,\ldots\nn\\
 F(z,0)&=&{z\01-z}\;,\qqd F(z,1)\;\,=\;\,\ln(1-z)\nn\\
 F(1,s)&=&\z(s)\qd\;\hbox{Riemann's zeta-function}\,.\label{FZs}\eea
Combining the results obtained so far it follows that
\beq J^{(4)}_l(t-t')\;\stackrel{t'\to t}{\sim}\;(t-t')^{-(d+l-2)}\;.
 \label{Jas}\eeq
This means that due to the behaviour of the integrand at the upper limit of
integration the convolutions in (\ref{Fki}) are logarithmically ($d=3$, $l=0$),
quadratically ($d=3$, $l=2$) or quartically ($d=3$, $l=4$) divergent,
if we suppose $r(t')$ and $I(t')$ to be smooth functions different from zero.\\
The divergent convolutions in (\ref{Fki}) represent non-local contributions to
the energy momentum expectation value (\ref{EIT6}). However the divergencies
emerge directly at the upper limit of integration and involve the functions
$r$ and $I$ (respectively the spacetime metric) only at the point $t'=t$.
Hence the divergencies are again of a local nature as claimed in
section \ref{FRWQFT}.\\
For the purpose of renormalization we have to isolate the divergencies from
the integrals (\ref{Fki}) in terms of $1/\e\,$-poles ($\e=3\!-\!d$). Taking into
account the behaviour (\ref{Jas}) of the convolution kernels it turns out that
this can be achieved by performing  $l\!+\!1$ integrations by parts. One
obtains divergent as well as finite boundary terms and finite
convolution integrals. At this point the variable $p$ appearing already in
(\ref{G1}) (being unused up to now) becomes meaningful: According to appendix
\ref{Hankel} our $G$-function satisfies the relations
\bea \biggl(\2\,\6_{t'}-\Bigl(p+1-{l\02}\Bigr)\biggr)G(t\!-\!t';p\!+\!1,l)
 &=&G(t\!-\!t';p,l)\label{dtG}\\
 \biggl(\2\,\6_{t'}-\Bigl(p+2-{l\02}\Bigr)\biggr)\tau_0^2(t\!-\!t')\,
 G(t\!-\!t';p\!+\!1,l)&=&\tau_0^2(t\!-\!t')\,G(t\!-\!t';p,l)\,,\nn\eea
which can be used to do the integrations by parts in (\ref{Fki}). The variable
$p$ then stands for the number of integrations by parts performed so far.
In this way we obtain for the integrals appearing in (\ref{Fki}):
\bea \lefteqn{\int\limits_0^t\!dt'\,r(t')\,G(t\!-\!t';0,0)\;=}\nn\\
 &&\2\,r(t')\,G(t\!-\!t';1,0)\bigg|_{t'=0}^{t'=t}\,-\int\limits_0^t\!dt'\,\Bigl(
 \2\,\9r(t')+r(t')\Bigr)\,G(t\!-\!t';1,0)\nn\\
 \lefteqn{\int\limits_0^t\!dt'\,I(t')\,G(t\!-\!t';0,2)\;=}\nn\\
 &&\2\biggl(I(t')\,G(t\!-\!t';1,2)-{\9I(t')\02}\,G(t\!-\!t';2,2)+
 \Bigl({\ddot I(t')\04}+{\9I(t')\02}\Bigr)G(t\!-\!t';3,2)\biggr)\bigg|_0^t\nn\\
 &&-\;\int\limits_0^t\!dt'\,\biggl({\dI(t')\08}+{3\04}\,\ddot I(t')+\9I(t')
 \biggr)\,G(t\!-\!t';3,2)\nn\\
\lefteqn{\int\limits_0^t\!dt'\,r(t')\,\tau_0^2(t\!-\!t')G(t\!-\!t';0,2)\;=}\nn\\
 &&\2\biggl(r(t')\,\tau_0^2\!G(t\!-\!t';1,2)-\Bigl({\9r(t')\02}+r(t')\Bigr)
 \,\tau_0^2\!G(t\!-\!t';2,2)\nn\\
 &&\hphantom{\2\biggl(}\hskip125pt+\;\Bigl({\ddot r(t')\04}+{3\02}\,\9r(t')
 +2\,r(t')\Bigr)\,\tau_0^2\!G(t\!-\!t';3,2)\biggr)\bigg|_0^t\nn\\
 &&-\;\int\limits_0^t\!dt'\,\biggl({\dr(t')\08}+{3\02}\,\ddot r(t')+
 {11\02}\,\9r(t')+6\,r(t')\biggr)\,\tau_0^2\!G(t\!-\!t';3,2)\nn\\
\lefteqn{\int\limits_0^t\!dt'\,I(t')\,\tau_0^2(t\!-\!t')G(t\!-\!t';0,4)\;=}\nn\\
 &&\2\biggl(I(t')\,\tau_0^2\!G(t\!-\!t';1,4)-{\9I(t')\02}\,
 \tau_0^2\!G(t\!-\!t';2,4)\nn\\
 &&\hphantom{\2\biggl(}+\;\Bigl({\ddot I(t')\04}+{\9I(t')\02}\Bigr)
 \,\tau_0^2\!G(t\!-\!t';3,4)-\Bigl({\dI(t')\08}+{3\04}\,\ddot I(t')
 +\9I(t')\Bigr)\,\tau_0^2\!G(t\!-\!t';4,4)\nn\\
 &&\hphantom{\2\biggl(}\hskip76pt+\;\Bigl({I^{\rm(iv)}(t')\016}+{3\04}\,\dI(t')+
 {11\04}\,\ddot I(t')+3\,\9I(t')\Bigr)\,\tau_0^2\!G(t\!-\!t';5,4)\biggr)
 \bigg|_0^t\nn\\
 &&-\;\int\limits_0^t\!dt'\,\biggl({I^{\rm(v)}(t')\032}+{5\08}\,I^{\rm(iv)}(t')
 +{35\08}\,\dI(t')+{25\02}\,\ddot I(t')+12\,\9I(t')\biggr)
 \,\tau_0^2\!G(t\!-\!t';5,4)\nn\\ \label{part}\eea
The convolutions we are left with in (\ref{part}) are convergent and finite.
The same holds for the lower boundary terms (proportional to $G(t;p,l)$) at
$t>0$. The upper boundary terms proportional to $G(0;p,l)$ are divergent of
the order $\s+1=d+l-2-p$, because they contain the functions
\bea \lefteqn{\vFd(\a_1,\a_2,\a_3,\a_4;\be_1,\be_2,\be_3;1)\;=
 \;\sum_{n=0}^{\infty}Q_n\;=\;\sum_nn^{\s}+\,\ldots}\nn\\
 &=&\z(-\s)+\8B_1\,\z(-\s\!+\!1)+\8B_2\,\z(-\s\!+\!2)+\8B_3\,\z(-\s\!+\!3)
 +\8B_4\,\z(-\s\!+\!4)\nn\\ &&+\;\hbox{finite terms}\;.\label{F2}\eea
In the second line above we have already given the analytic continuation
of the divergent part of the series $\vFd(\ldots;\ldots;1)$ using Riemann's
zeta-function. We have taken into account
as many terms as necessary for the ``most divergent'' case
$G(0;1,4)$ with $\s=d=3-\e$.\\
The zeta-function has exactly one simple pole at $s=1\,$:
\beq \z(s)\;=\;{1\0s-1}+\g+\sum_{n=1}^{\infty}\g_n(s-1)^n\;, \eeq
where $\g$ is the Euler-Mascheroni constant (see \cite{HTI}). This pole
becomes an $1/\e\,$-pole in (\ref{F2}) and represents the divergencies in the
typical manner for dimensional regularization. In order to separate them
from the finite part we rewrite (\ref{F2}) in the form
\beq \vFd(\a_1,\ldots,\be_3;1)\;=\;
 \vFe(\a_1,\ldots,\be_3)+\sum_{m=0}^M
 \8B_m(\a_1,\ldots,\be_3,1)\,\z(m-\s)\,,\label{F3}\eeq
where we have defined the convergent series $\vFe$ in the following way:
\bea \lefteqn{\vFe(\a_1,\ldots,\be_3)\;:=\;}\label{Ftil}\\
 &&Q_0(\a_1,\ldots,\be_3)
 +\sum_{n=1}^{\infty}\biggl(Q_n(\a_1,\ldots,\be_3)-
 \sum_{m=0}^M\8B_m(\a_1,\ldots,\be_3,1)\,n^{\s-m}\biggr)\nn\eea
The number $M+1$ of terms to be subtracted from every term of the series is
determined by $\a_1\ldots\be_3$ in such a way that the series
(\ref{Ftil}) is just convergent: $M$ is the biggest integer less or equal to
$\s+1$, and in our case for $d=3$ we have $M=l-p+1$.\\
In (\ref{Ftil}) $d=3$ may already be substituted. The terms of the series
behave like $1/n^2$ for large $n$, hence a truncation of the series at $n=N$
(for example for an approximate numerical calculation) will lead to an error
of the order of $1/N$.\\
The only term in (\ref{F3}), for which the regularization $d=3-\e$ has to be
retained until renormalization, is the $m\!=\!M\,$-term in the zeta-function
sum containing the $1/\e\,$-pole.\\
Now we define a finite function $\8G(p,l)$ in the same manner as previously
$G(t;p,l)$, excepted that $t=0$ and the $\vFd\,$'s in the definition (\ref{G1})
have to be replaced by the corresponding $\vFe\,$'s.\\
The functions $G(0;p,l)\,$ which are needed for the upper boundary terms
in (\ref{part}) can then be expressed in terms of $\8G(p,l)$ and
zeta-functions. Using (\ref{G1}), (\ref{F3}), (\ref{Btil}), (\ref{Bbar}) und (\ref{Bern})
we obtain in the limit $\e\to0$:
\bean G(0;1,0)\!&=&\!\8G(1,0)+\z(1+\e)\nn\\
 G(0;1,2)\!&=&\!\8G(1,2)+\z(-1)+(1+i\nu\cot\pi\nu)\,\z(0)\nn\\
 &&+\,\biggl({3\02}\Bigl(\4\!-\!\nu^2\Bigr)+\e\Bigl({\nu^2\04\sin^2\pi\nu}
 -{29\048}-i\nu\cot\pi\nu\Bigr)\biggr)\z(1+\e)\nn\\
 G(0;2,2)\!&=&\!\8G(2,2)+\z(0)+\biggl(\2-\e\Bigl({3\04}+i\nu\cot\pi\nu\Bigr)
 \biggr)\z(1+\e)\nn\\
 G(0;3,2)\!&=&\!\8G(3,2)+\z(1+\e)\nn\\
 G(0;1,4)\!&=&\!\8G(1,4)+\z(-3)+\biggl({9\02}+3i\nu\cot\pi\nu\biggr)\z(-2)\nn\\
 &&+\,\biggl({57\08}+{\nu^2\02}-{3\nu^2\02\sin^2\pi\nu}+9i\nu\cot\pi\nu\biggr)
 \z(-1)\nn\\
 &&+\,\biggl({29\08}+2\nu^2-{9\nu^2\04\sin^2\pi\nu}
 +i\Bigl({57\08}\nu\!-\!{3\02}\nu^3\Bigr)\cot\pi\nu\biggr)\z(0)\nn\\
 &&+\,\biggl({15\04}\Bigl(\4\!-\!\nu^2\Bigr)+{15\08}\Bigl(\4\!-\!\nu^2\Bigr)^2
 \nn\\ &&\hphantom{+\,\biggl(}-\e\Bigl({407\0240}+
 {47\096}\nu^2+{\nu^4\!-\!57\nu^2\!/16\02\sin^2\pi\nu}-i\Bigl(\nu^3\!-\!
 {29\08}\nu\Bigr)\cot\pi\nu\Bigr)\biggr)\z(1+\e)\nn\\
\eean\newpage 
\bea
 G(0;2,4)\!&=&\!\8G(2,4)+\z(-2)+(3+2i\nu\cot\pi\nu)\,\z(-1)\nn\\
 &&+\,\biggl({21\08}-{3\02}\nu^2-{\nu^2\02\sin^2\pi\nu}+3i\nu\cot\pi\nu\biggr)
 \z(0)+\biggl(-{5\04}\Bigl(\4\!-\!\nu^2\Bigr)\nn\\
 &&\hphantom{+\,\biggl(}-\e\Bigl({49\032}-{11\08}\nu^2-{3\nu^2\04\sin^2\pi\nu}
 +i\Bigl({21\08}\nu\!-\!{13\06}\nu^3\Bigr)\cot\pi\nu\Bigr)\biggr)\z(1+\e)\nn\\
 G(0;3,4)\!&=&\!\8G(3,4)+\z(-1)+\biggl({5\02}+i\nu\cot\pi\nu\biggr)\z(0)\nn\\
 &&+\,\biggl({11\08}-{5\02}\nu^2-\e\Bigl({83\048}-{\nu^2\04\sin^2\pi\nu}
 +{5\02}\,i\nu\cot\pi\nu\Bigr)\biggr)\z(1+\e)\nn\\
 G(0;4,4)\!&=&\!\8G(4,4)+\z(0)+\biggl(3-\e\Bigl({3\04}+i\nu\cot\pi\nu\Bigr)
 \biggr)\z(1+\e)\nn\\
 G(0;5,4)\!&=&\!\8G(5,4)+\z(1+\e) \label{G2}\eea
The zeta-functions have the following values
\bea &&\z(1+\e)\,=\,{1\0\e}+\g\;,\qd \z(-n)\,=\,{(-)^nB_{n+1}\0n+1}\qd
 n\in\hbox to 1.5pt{$I$\hss}N_0 \nn\\
 \then&&\z(0)\,=\,-\2\,,\qd\z(-1)\,=\,-{1\012}\,,\qd\z(-2)\,=\,0\,,\qd
 \z(-3)\,=\,{1\0120}\,.\hskip12pt\label{zeta}\eea
With the help of (\ref{Fk})--(\ref{dGam}), (\ref{Int0}),
(\ref{Fki}), (\ref{Fkii}), (\ref{J4l1}), (\ref{part}), (\ref{G2}) and (\ref{zeta})
we are in the position to specify the rest of the integrals appearing in
(\ref{EIT6}):\newpage
\bea \lefteqn{\int\!\8d\8k\,\de\biggl({1\02\,\re A}\biggr)\;=:\;\de J^{(2)}_0
 \;=\;\de J^{(2)}_{0\,\rm div}+\de J^{(2)}_{0\,\rm fin}}\nn\\
 \lefteqn{\de J^{(2)}_{0\,\rm div}\;=\;{1\08\pi^2}\,\xi_c\de R(t)
 \biggl({1\0\e}+{\g\02}+1+\2\ln\pi\biggr)}\nn\\
 \lefteqn{\de J^{(2)}_{0\,\rm fin}\;=\;{1\08\pi^2}\biggl(
 -{5\04}\,\9I(t)+I(t)\,\Bigl({57\024}-{\nu^2\02\sin^2\pi\nu}\Bigr)}\nn\\
 &&-\,I(t)\,\8m_0^2\Bigl({5\02}+3\g-3\ln2+{3\02}\,\psi({\ts{3\02}\!+\!\nu})
 +{3\02}\,\psi({\ts{3\02}\!-\!\nu})\Bigr)\nn\\
 &&+\,r(t)\,\8G_R(1,0)-2I(t)\,\8G_R(1,2)+\9I(t)\,\8G_R(2,2)
 -\Bigl({\ddot I(t)\02}+\9I(t)\Bigr)\,\8G_R(3,2)\nn\\
 &&-\,r(0)\,G_R(t;1,0)+2I(0)\,G_R(t;1,2)-\9I(0)\,G_R(t;2,2)\nn\\
 &&+\,\Bigl({\ddot I(0)\02}+\9I(0)\Bigr)\,G_R(t;3,2)
 -\int\limits_0^t\!dt'\,(\9r(t')+2r(t'))\,G_R(t\!-\!t';1,0)\nn\\
 &&+\,\int\limits_0^t\!dt'\,\Bigl({\dI(t')\02}+3\ddot I(t')+4\9I(t')\Bigr)\,
 G_R(t\!-\!t';3,2)\nn\\
 &&+\,2\sum_{n=-3}^{+1}\de\G_n\,G_R(t;0,n)-{\pi^2\02}\,\re
 \int\limits_0^{\infty}\!dy\,y^2{H^{(1)}_{\nu}}^2\!(y)\,
 {H^{(2)}_{\nu}}^2\!(ya_0(t))\,\de\G^{(ii)}(ya_0(t))\biggr)\nn\\
 \label{T1}\eea
The abbreviations $\8G_R(p,l):=\re\,\8G(p,l)$ and $G_R(t;p,l):=\re\,G(t;p,l)$
have been used.\\
Apart from the $1/\e\,$-pole in (\ref{T1}) and (\ref{T2}), which will be removed
by renormalization, the functions $G_R(t;p,l)$ are divergent in the limit
$t\to0$. Using explicitly the asymptotic expansion (\ref{Fas}) and the
coefficients $\de\G_n$ (\ref{dGamn}) it turns out that the
$t\!\to\!0\,$-divergencies of the lower boundary terms in (\ref{part}) cancel
the $t\!\to\!0\,$-divergencies from our $F^{(ii)}_k$ resp. $\de\G(k,0)$
as it has to be.
It is for this reason that we need the $\de\G_n\,$-terms and the comparison
with the adiabatic vacuum in section \ref{adia}. In the same way the
finiteness of the first and second time derivatives of
$\de J^{(2)}_{0\,\rm fin}$ needed in (\ref{EIT6}) has been checked for $t\to0$.
\newpage
\bea \lefteqn{\int\!\8d\8k\,\de\biggl({a^{-2}k^2\02\,\re A}\biggr)\;=:\;
 \de J^{(2)}_2\;=\;\de J^{(2)}_{2\,\rm div}+\de J^{(2)}_{2\,\rm fin}}\nn\\
 \lefteqn{\de J^{(2)}_{2\,\rm div}\;=\;{1\08\pi^2}
 \biggl({\8m_0^2\02}\,\ddot I(t)+3\8m_0^2\9I(t)-{3\02}\,(\8m_0^2\!+\!1)\,\xi_c
 \de R(t)-{5\04}\,\xi_c\de\9R(t)-\4\,\xi_c\de\ddot R(t)\biggr)}\nn\\
 \lefteqn{\hphantom{\de J^{(2)}_{2\,\rm div}\;=\;{1\08\pi^2}}
 \cdot\biggl({1\0\e}+{\g\02}+1+\2\ln\pi\biggr)}\nn\\
 \lefteqn{\de J^{(2)}_{2\,\rm fin}\;=\;{1\08\pi^2}\biggl(
 -I(t)\Bigl({131\016}+{49\016}\nu^2+{\nu^4\!-\!97\nu^2\!/16\0\sin^2\pi\nu}
 \Bigr)+\9I(t)\Bigl({5\02}-{17\08}\nu^2-{7\,\nu^2\08\sin^2\pi\nu}\Bigr)}\nn\\
 &&-\,\xi_c\de R(t)\Bigl({3\048}+{\nu^2\04\sin^2\pi\nu}\Bigr)
 -{5\08}\,\xi_c\de\9R(t)\nn\\
 &&+\,{15\04}\,I(t)\,\8m_0^2\Bigl({3\02}+(m^2\!-\!\xi R_0)\Bigl({47\060}+\g
 -\ln2+\2\,\psi({\ts{3\02}\!+\!\nu})+\2\,\psi({\ts{3\02}\!-\!\nu})\Bigr)\Bigr)\nn\\
 &&-\,r(t)\,\8G_R(1,2)+\Bigl({\9r(t)\02}+r(t)\Bigr)\,\8G_R(2,2)
 -\Bigl({\ddot r(t)\04}+{3\02}\,\9r(t)+2r(t)\Bigr)\,\8G_R(3,2)\nn\\
 &&+\,2I(t)\,\8G_R(1,4)-\9I(t)\,\8G_R(2,4)-\Bigl({\dI(t)\04}+{3\02}\,\ddot I(t)
 +2\9I(t)\Bigr)\,\8G_R(4,4)\nn\\
 &&+\,\Bigl({\ddot I(t)\02}+\9I(t)\Bigr)\,\8G_R(3,4)+
 \Bigl({I^{\rm(iv)}(t)\08}+{3\02}\,\dI(t)+{11\02}\,\ddot I(t)+6\9I(t)\Bigr)
 \,\8G_R(5,4)\nn\\
 &&+\,r(0)\,\tau_0^2\!G_R(t;1,2)-\Bigl({\9r(0)\02}+r(0)\Bigr)\,\tau_0^2\!
 G_R(t;2,2)\nn\\
 &&+\Bigl({\ddot r(0)\04}+{3\02}\,\9r(0)+2r(0)\Bigr)\tau_0^2\!G_R(t;3,2)
 -2I(0)\,\tau_0^2\!G_R(t;1,4)+\9I(0)\,\tau_0^2\!G_R(t;2,4)\nn\\
 &&-\,\Bigl({\ddot I(0)\02}+\9I(0)\Bigr)\,\tau_0^2\!G_R(t;3,4)
 +\Bigl({\dI(0)\04}+{3\02}\,\ddot I(0)+2\9I(0)\Bigr)\,\tau_0^2\!G_R(t;4,4)\nn\\
 &&-\,\Bigl({I^{\rm(iv)}(0)\08}+{3\02}\,\dI(0)+{11\02}\,\ddot I(0)+6\9I(0)\Bigr)
 \,\tau_0^2\!G_R(t;5,4)\nn\\
 &&+\,\int\limits_0^t\!dt'\,\Bigl({\dr(t')\04}+3\,\ddot r(t')+11\,\9r(t')
 +12\,r(t')\Bigr)\,\tau_0^2\!G_R(t\!-\!t';3,2)\nn\\
 &&-\,\int\limits_0^t\!dt'\,\Bigl({I^{\rm(v)}(t')\08}+{5\02}\,I^{\rm(iv)}(t')
 +{35\02}\,\dI(t')+50\,\ddot I(t')+48\,\9I(t')\Bigr)\,
 \tau_0^2\!G_R(t\!-\!t';5,4)\nn\\
 &&-\,2\!\sum_{n=-3}^{+1}\!\de\G_n\tau_0^2\!G_R(t;0,n\!+\!2)-{\pi^2\02}\,\re
 \int\limits_0^{\infty}\!dy\,y^4{H^{(1)}_{\nu}}^2\!(y)\,
 {H^{(2)}_{\nu}}^2\!(ya_0(t))\,\de\G^{(ii)}(ya_0(t))\biggr)\!\nn\\
 \label{T2}\eea

\subsection{Renormalization}
Our renormalization scheme consists in the subtraction of De~Witt-Schwinger
terms as was explained in section \ref{FRWQFT}. Again we linearize with
respect to the deviation from the de~Sitter -- Bunch-Davies system. With the
aid of appendix \ref{geoT} equation (\ref{TDS2}) leads to:
\bea \de\r_{\rm DS\,div}\!&=&\!{1\016\pi^2}\biggl({1\0\e}
 -\2\Bigl(\g+\ln{m^2\04\pi}\Bigr)\biggr)\nn\\
 &&\cdot\biggl({-4\,m^2\0d\!-\!1}\,(\xi\!-\!{\ts{1\06}})\,\de G_{00}
 +{1\090}\Bigl(\de H_{00}-\de\,{}^{(2)}\!H_{00}\Bigr)+(\xi\!-\!{\ts{1\06}})^2
 \de\,{}^{(1)}\!H_{00}\biggr)\nn\\
 &=&\!{1\016\pi^2}\biggl({1\0\e}-\2\Bigl(\g+\ln{m^2\04\pi}\Bigr)\biggr)
 \biggl({-4\,m^2\0d\!-\!1}\,(\xi\!-\!{\ts{1\06}})\,d(d\!-\!1)\,\9I\nn\\
 &&\qqd+\,{1\090}\,d(d\!-\!3)\Bigl(\dI+d\,\ddot I+2(d\!-\!2)\,\9I\Bigr)\nn\\
 &&\qqd-\,(\xi\!-\!{\ts{1\06}})^2d^2\Bigl(4\,\dI+4d\,\ddot I+2(d\!+\!1)
 (d\!-\!3)\,\9I\Bigr)\biggr)\label{rDS}\\
 \de p_{\rm DS\,div}\!&=&\!{1\016\pi^2}\biggl({1\0\e}-\2\Bigl(\g+\ln{m^2\04\pi}
 \Bigr)\biggr)\biggl({-4\,m^2\0d\!-\!1}\,(\xi\!-\!{\ts{1\06}})\,(1\!-\!d)
 \Bigl(\ddot I+d\,\9I\Bigr)\nn\\
 &&\qqd+\,{1\090}\,(3\!-\!d)\Bigl(I^{\rm(iv)}+2d\,\dI+(d^2\!+\!2d\!-\!4)\,
 \ddot I+2d(d\!-\!2)\,\9I\Bigr)\nn\\
 &&\qqd+\,(\xi\!-\!{\ts{1\06}})^2d\Bigl(4\,I^{\rm(iv)}+8d\,\dI+
 (6d^2\!-\!4d\!-\!6)\,\ddot I+2d(d\!-\!3)(d\!+\!1)\,\9I\Bigr)\biggr)\nn\eea
Together with (\ref{EIT6}), (\ref{T1}) and (\ref{T2}) we finally obtain the
components of the renormalized energy momentum tensor expectation value:
\newpage
\bea \de\r_{\rm ren}\!&=&\!\de\r-\de\r_{\rm DS\,div}\nn\\
 &=&\!\xi\,(\de R_{00}-\de R)\,J^{(2)}_0+(m^2+\xi\,d^2+d\,({\ts\xi\!+\!\4})\,
 \6_t+{\ts\4}\,\6_t^2)\,(\de J^{(2)}_{0\,\rm div}+\de J^{(2)}_{0\,\rm fin})\nn\\
 &&+\,\de J^{(2)}_{2\,\rm div}+\de J^{(2)}_{2\,\rm fin}-\de\r_{\rm DS\,div}\nn\\
 &=&\!{1\08\pi^2}\biggl({m^2\04}\,\ddot I+m^2(3\,\xi_c\!+\!{\ts{7\03}})\,\9I
 +2\xi_c\Bigl(\dI+3\ddot I\Bigr)+{27\02}\,\xi_c\ddot I+12\,\xi_c(3\xi_c\!+\!5)
 \,\9I\nn\\
 &&+\,{1\060}\Bigl(\dI+3\ddot I+2\9I\Bigr)-36\,\xi_c^2\9I\nn\\
 &&+\,{\8m_0^2\02}\,\xi\,(\de R_{00}-\de R)\Bigl({\ts\psi({3\02}\!+\!\nu)
 +\psi({3\02}\!-\!\nu)}-1-\ln m^2\Bigr)\nn\\
 &&+\,(m^2+9\,\xi+3\,({\ts\xi\!+\!\4})\,\6_t+{\ts\4}\,\6_t^2)\Bigl(\xi_c
 \de R\Bigl(\g+1+\2\ln{m^2\04}\Bigr)+8\pi^2\de J^{(2)}_{0\,\rm fin}\Bigr)\nn\\
 &&+\Bigl({\8m_0^2\02}\ddot I+3\8m_0^2\9I-{3\02}\,(\8m_0^2\!+\!1)\,\xi_c\de R
 -{5\04}\,\xi_c\de\9R-\4\,\xi_c\de\ddot R\Bigr)\Bigl(\g+1+\2\ln{m^2\04}\Bigr)
 \nn\\ &&+\,8\pi^2\de J^{(2)}_{2\,\rm fin}\,\biggr) \label{ex1}\eea
\bea \lefteqn{\de p_{\rm ren}\;=\;\de p-\de p_{\rm DS\,div}}\nn\\
 &=&\!{1\08\pi^2}\biggl(m^2\Bigl({5\036}\,\ddot I+\9I\Bigr)+m^2\xi_c\9I
 +12\,\xi_c^2\Bigl(\ddot I+4\9I\Bigr)+\xi_c\Bigl({I^{\rm(iv)}\03}+3\,\dI
 +{21\02}\,\ddot I+20\,\9I\Bigr)\nn\\
 &&-\,{1\0180}\Bigl(I^{\rm(iv)}+6\,\dI+11\,\ddot I+6\,\9I\Bigr)\nn\\
 &&+\,{\8m_0^2\02}\,\xi\,\de(a^{-2}R_{11})\Bigl({\ts\psi({3\02}\!+\!\nu)
 +\psi({3\02}\!-\!\nu)}-1-\ln m^2\Bigr)\nn\\
 &&+\,(\xi\,a^{-2}R_{11}+({\ts{3\04}}\!-\!2\xi)\,\6_t+({\ts\4}\!-\!\xi)\,\6_t^2)
 \Bigl(\xi_c
 \de R\Bigl(\g+1+\2\ln{m^2\04}\Bigr)+8\pi^2\de J^{(2)}_{0\,\rm fin}\Bigr)\nn\\
 &&+\,{1\03}\Bigl({\8m_0^2\02}\ddot I+3\8m_0^2\9I-{3\02}\,
 (\8m_0^2\!+\!1)\,\xi_c\de R-{5\04}\,\xi_c\de\9R
 -\4\,\xi_c\de\ddot R\Bigr)\Bigl(\g+1+\2\ln{m^2\04}\Bigr)\nn\\
 &&+\,{8\pi^2\03}\,\de J^{(2)}_{2\,\rm fin}\,\biggr) \label{ex2}\eea
According to $d\!=\!3\,$, $\xi_c=\xi-{1\06}$.\\
All divergencies have cancelled and the final results (\ref{ex1}) and (\ref{ex2})
are finite. This fact can be regarded a non-trivial check on the calculation.\\
Contrary to the other terms the $\de J^{(2)}_{l\,\rm fin}\,$'s are non-local
functionals of $r(t)$ and $I(t)$, because they contain convolution integrals
as well as the initial data at $t=0$.\\
The $G$-functions appearing in the $\de J^{(2)}_{l\,\rm fin}\,$'s in (\ref{T1})
and (\ref{T2}) are defined as convergent series and are therefore well suited
for a numerical computation.

\expandafter\ifx\csname ok\endcsname\relax
   \end{document}\fi
\expandafter\ifx\csname qd\endcsname\relax
   \documentstyle[12pt,twoside]{article}\let\qd=\quad \let\qqd=\qquad \def\qqqd{\qquad\qquad}

\let\a=\alpha \let\be=\beta \let\g=\gamma \let\de=\delta
\let\e=\varepsilon \let\z=\zeta \let\h=\eta \let\th=\theta
\let\dh=\vartheta \let\k=\kappa \let\la=\lambda
\let\r=\rho \let\s=\sigma
\let\om=\omega
\let\ph=\varphi
\let\Om=\Omega \let\Si=\Sigma \let\Th=\Theta
\let\La=\Lambda \let\G=\Gamma \let\D=\Delta

\def\0{\over } \def\1{\vec } \def\2{{1\over2}} \def\4{{1\over4}}
\def\5{\bar } \def\6{\partial } \def\7#1{{#1}\llap{/}}
\def\8{\tilde } \def\9{\dot }

\def\ul#1{$\underline{\hbox{#1}}$} \def\llp{\hbox to 0pt{\hss /\hskip1.5pt}}
\def\so{\supset\hbox to 0pt{\hss $\displaystyle -$\hskip1pt}}

\def\ts{\textstyle } \def\ds{\displaystyle }
\def\<{\langle } \def\>{\rangle } \def\lb{\left} \def\rb{\right}
\let\lra=\leftrightarrow \let\LRA=\Leftrightarrow
\let\then=\Rightarrow \def\triang{{\scriptstyle \Delta}}
\def\trian{{\scriptscriptstyle \Delta}}
\def\re{{\rm Re}} \def\im{{\rm Im}} \def\ord{{\cal O}}

\let\ap=\approx \let\eq=\equiv \let\ex=\times \let\hc=\dagger
      \let\ti=\tilde \let\bl=\biggl \let\br=\biggr
\let\bi=\choose \let\at=\atop \let\mat=\pmatrix

\let\nn=\nonumber
\def\bea{\begin{eqnarray}} \def\eea{\end{eqnarray}}
\def\bean{\begin{eqnarray*}} \def\eean{\end{eqnarray*}}
\def\beq{\begin{equation}} \def\eeq{\end{equation}}
\def\barr{\begin{array}} \def\earr{\end{array}}

\newcommand{\deflabel}[1]{#1\hfil}%
\newenvironment{deflist}[1]%
{\begin{list}{}%
{\settowidth{\labelwidth}{#1}%
\setlength{\leftmargin}{\labelwidth}\addtolength{\leftmargin}{\labelsep}%
\renewcommand{\makelabel}{\deflabel}}}%
{\end{list}}%
\fi
\expandafter\ifx\csname ok\endcsname\relax
   \def\re{{\rm Re}} \def\im{{\rm Im}}
   {\catcode`@=11 \@addtoreset{equation}{section}\@addtoreset{figure}{section}}
   \def\theequation{\thesection.\arabic{equation}}
   \def\thefigure{\thesection.\arabic{figure}}
   \addtolength{\topmargin}{-48pt}
   \addtolength{\textheight}{90pt}
   \addtolength{\evensidemargin}{-31pt}
   \addtolength{\oddsidemargin}{-12pt}
   \addtolength{\textwidth}{30pt}
   \addtolength{\footskip}{21pt}
   \setlength{\parindent}{0pt}\frenchspacing
   \def\vFd{{}_4\5F_3}
\def\dI{I\raise10.1pt\hbox to 0pt{\hss.\hskip-1.44pt.\hskip-1.44pt.\hskip-1.8pt}}
   \includeonly{}
   \begin{document}
   
\fi
\section{The linearized Einstein equations}\label{solv}
Restricting on FRW spacetimes the semiclassical Einstein equations
\beq R_{\mu\nu}-\2\,g_{\mu\nu}R+\La\,g_{\mu\nu}\;=\;8\pi G_N\,
  \<T_{\mu\nu}\>_{\rm ren} \label{Ein1}\eeq
are containing two independent components.
They read (remember $H(t):=\9a/a=H_0+\9I(t)$)
\bea 3\,H^2+\La&=&8\pi G_N\,\r \label{Ein2}\\
 2\,\9H+3\,H^2+\La&=&-8\pi G_N\,p\;.\label{Ein3}\eea
The index ``ren'' at $\r$ and $p$ will be suppressed and the Hubble constant
$H_0$ will be explicitly written out in this section.\\
If the sources $\r$ and $p$ are specified in advance equation (\ref{Ein2}) (the
00-component of (\ref{Ein1})) is no dynamical equation of motion but a
constraint on the initial data. It plays the role of the Poisson equation
in electrodynamics. Here instead the sources are themselves functionals of
the metric and are reacting on its changes. Therefore equation (\ref{Ein2})
is a dynamical equation in the present case.\\
Neither the dimensional regularization nor our renormalization scheme are
spoiling the covariant energy momentum conservation:
\beq D_{\mu}\<T^{\mu0}\>_{\rm ren}\;=\;\9{\r}+3H(\r+p)\;=\;0\label{EIE2}\eeq
On that account the equations (\ref{Ein2}) and (\ref{Ein3}) are not independent:
Every solution of (\ref{Ein2}) is also a solution to (\ref{Ein3}). Therefore,
only (\ref{Ein2}) is considered in the following.
In its linearized form it reads:
\beq 6\,{\9I(t)\0H_0}\;=\;{8\pi G_N\0H_0^2}\,\de\r[I;t] \label{Ein4}\eeq
Due to the convolutions contained in $\de\r$ this is a linear
integro-differential equation and may be conveniently solved by
Laplace transformation.

\subsection{Laplace transformation}\label{Laplace}
The Laplace transform of a function $f(t)$ will be denoted by
${\cal L}[f;s]$ or $\widehat{f}(s)$:
\beq {\cal L}[f;s]\;:=\;\widehat{f}(s)\;:=\;\int\limits_0^{\infty}\!dt\,
 f(t)\,e^{-st} \eeq
Utilizing
\[ \widehat{\9f}(s)\,=\,s\widehat{f}(s)-f(0)\,,\qqd\widehat{f^{(n)}}(s)\,=\,
 s^n\widehat{f}(s)-s^{n-1}f(0)-\ldots-f^{(n-1)}(0)\,,\]
\beq {\cal L}\biggl[\int\limits_0^t\!dt'\,f(t')\,g(t\!-\!t')\,;s\biggr]\,=\,
 \widehat{f}(s)\cdot\widehat{g}(s)\,\eeq
one is able to compute the Laplace transform $\widehat{\de\r}(s)$ of
$\de\r$ (\ref{ex1}). For the $\de J^{(2)}_{\,\rm fin}\,$'s we need the
transforms $\widehat{G_R}(s;p,l)$ and $\widehat{\tau_0^2\!G_R}(s;p,l)$ of the
functions $G_R(t;p,l)$ and $\tau_0^2(t)G_R(t;p,l)\,$, which were defined in
(\ref{G1}) in terms of three generalized hypergeometric series
$\vFd(\ldots;\ldots;\tau_0^2(t))\,$. Their Laplace transforms are again
hypergeometric series:
\bea {\cal L}\Bigl[\tau_0^{2\a}\vFd(\ldots;\ldots;\tau_0^2(t))\,;s\Bigr]&=&
{\cal L}\Bigl[e^{-2\a H_0t}\sum_{n=0}^{\infty}Q_n(\ldots;\ldots)\,e^{-2H_0tn}\,;
 s\Bigr]\nn\\
 &=&\sum_{n=0}^{\infty}{Q_n(\ldots;\ldots)\0s+2H_0n+2\a H_0}\label{LFb}\\
 &=&{1\02H_0}\,{}_5\5F_4\Bigl({s\02H_0}+\a,\dots;{s\02H_0}+\a+1,\ldots;1)
 \nn\eea
The particular functions $\vFd(\ldots;\ldots;\tau_0^2(t))$ in the
$\de J^{(2)}_{\,\rm fin}\,$'s (\ref{T1}) and (\ref{T2}) are singular for $t\to0$,
so that the corresponding series (\ref{LFb}) do not converge. Therefore all
$\vFd$-series in $\de J^{(2)}_{0\,\rm fin}$ respectively
$\de J^{(2)}_{2\,\rm fin}$ have to be added term by term before summing up the
series. The complete $\de J^{(2)}_{\,\rm fin}\,$'s are well behaved for $t\to0$
(see the comments after (\ref{T1})), and the term by term addition of the
series (\ref{LFb}) will be convergent.\\
The series $\widehat{G_R}(s;1,0)$, $\widehat{G_R}(s;3,2)$ and
$\widehat{G_R}(s;5,4)$ appearing in the convolutions are convergent by
themselves. For large $s$ they are of the order of $s^{-1}(1+\ln s)\,$.\\
The Laplace transform $\widehat{\de\r}(s)$ may be cast into the form
\beq \widehat{\de\r}(s)\;=:\;{H_0^4\08\pi^2}\biggl(\de\r_{\widehat{I}}
 \Bigl({s\0H_0}\Bigr)\,\widehat{I}(s)+\sum_{n=0}^2f_n\Bigl({s\0H_0}\Bigr)\,
 {I^{(n)}(0)\0H_0^{n+1}}+{1\0H_0}\,g\Bigl({s\0H_0}\Bigr)\biggr)\label{drh}\eeq
with certain functions $\de\r_{\widehat{I}}(s/H_0)\,$, $f_n(s/H_0)$ and
$g(s/H_0)\,$. The function $g$ contains the $\de\G(k,0)\,$- resp. $\de\G_n\,$-
and $\de\G^{(ii)}(k)\,$-contributions.\\
The explicit calculation shows\footnote{This has to prove true since (\ref{ex1})
results from (\ref{EIT6}) and (\ref{rDS}).} that the $\dI(0)\,$- and the
$I^{\rm(iv)}(0)\,$-terms drop out in (\ref{drh}). Only up to second derivatives
of $I(t)$ at time $t=0$ are appearing in $\widehat{\de\r}(s)$ and thereby in
(\ref{Ein5}) (apart from the $\de\G_n\,$, see below). This means that we have to
specify the initial data $\de\G(k,0)$, $I(0)$, $\9I(0)$ and $\ddot I(0)$ in
order to fix the solution of (\ref{Ein4}). Regarding $\de\G(k,0)$
we only have $\de\G^{(ii)}(k)$ at our disposal (see (\ref{dGam}) and afterwards),
whereas (\ref{dGamn}) has to be used for the $\de\G_n\,$.
The values of $\dI(0)$ and $I^{\rm(iv)}(0)$ required in (\ref{dGamn}) may be
obtained from equation (\ref{Ein4}) and its derivative at time $t=0$.
However, due to the smallness of Newton's constant it is more advantageous to
specify $\dI(0)$ instead of $\9I(0)$ and to determine $\9I(0)$ from (\ref{Ein4})
at $t=0$. Then it can be easily granted that all $I^{(n)}$ are small at $t=0$
and that the linearization is justified.\\
We want to save the explicit statement of the lengthy and in the following
unessential functions $f_n(s)$ and $g(s)$, but at least
$\de\r_{\widehat{I}}(s)$ should be given. The expression has been left in a
somewhat uncompactified shape, because there is presumably not much
shortening to gain:\newpage
\bea \lefteqn{\de\r_{\widehat{I}}(s)\;=\;
 {m^2\04H_0^2}\,s^2+{m^2\0H_0^2}\,(3\xi_c\!+\!{\ts{7\03}})\,s
 +2\xi_c\Bigl(s^3+3s^2\Bigr)+{27\02}\,\xi_cs^2+12\,\xi_c(3\xi_c\!+\!5)\,s}\nn\\
 &&+\,{1\060}\Bigl(s^3+3s^2+2s\Bigr)-36\,\xi_c^2s\nn\\
 &&+\,{3\8m_0^2\02H_0^2}\,\xi\,\Bigl(s^2+6s\Bigr)\Bigl({\ts\psi({3\02}\!+\!\nu)
 +\psi({3\02}\!-\!\nu)}-1-\ln{m^2\0H_0^2}\Bigr)\nn\\
 &&+\,\Bigl({m^2\0H_0^2}+9\,\xi+3\,({\ts\xi\!+\!\4})\,s+{\ts\4}\,s^2\Bigr)
 \biggl(-6\xi_c(s^2+4s)\Bigl(\g+1+\2\ln{m^2\04H_0^2}\Bigr)\nn\\
 &&\qd-\,{5\04}\,s+{57\024}-{\nu^2\02\sin^2\pi\nu}
 -{\8m_0^2\0H_0^2}\Bigl({5\02}+3\g-3\ln2+{3\02}\,\psi({\ts{3\02}\!+\!\nu})
 +{3\02}\,\psi({\ts{3\02}\!-\!\nu})\Bigr)\nn\\
 &&\qd+\,\Bigl({s^2\02}+{s\02}-6\xi_c(s^2+4s)\Bigr)\,\8G_R(1,0)\nn\\
 &&\qd-\,2\,\8G_R(1,2)+s\,\8G_R(2,2)-\Bigl({s^2\02}+s\Bigr)\,\8G_R(3,2)\nn\\
 &&\qd-\,(s+2)\Bigl({s^2\02}+{s\02}-6\xi_c(s^2+4s)\Bigr)\,\widehat{G_R}(s;1,0)
 +\Bigl({s^3\02}+3s^2+4s\Bigr)\,\widehat{G_R}(s;3,2)\biggr)\nn\\
 &&+\biggl({\8m_0^2\02H_0^2}\,s^2+3\,{\8m_0^2\0H_0^2}\,s+6\xi_c(s^2+4s)\Bigl(
 {3\02}\,\Bigl({\8m_0^2\0H_0^2}+1\Bigr)
 +{5\04}\,s+\4\,s^2\Bigr)\biggr)\nn\\
 &&\qqd\cdot\,\biggl(\g+1+\2\ln{m^2\04H_0^2}\biggr)
 -{131\016}-{49\016}\nu^2-{\nu^4\!-\!97\nu^2\!/16\0\sin^2\pi\nu}\nn\\
 &&+\Bigl({5\02}-{17\08}\nu^2-{7\,\nu^2\08\sin^2\pi\nu}\Bigr)\,s
 +\Bigl({3\08}+{3\,\nu^2\02\sin^2\pi\nu}+{15\04}\,s\Bigr)\,\xi_c(s^2+4s)\nn\\
 &&+\,{15\8m_0^2\04H_0^2}\biggl({3\02}+\Bigl({\8m_0^2\0H_0^2}+2\Bigr)
 \Bigl({47\060}+\g-\ln2+\2\,\psi({\ts{3\02}\!+\!\nu})
 +\2\,\psi({\ts{3\02}\!-\!\nu})\Bigr)\biggr)\nn\\
 &&+\Bigl({s^2\02}+{s\02}-6\xi_c(s^2+4s)\Bigr)
 \biggl(-\8G_R(1,2)+\Bigl({s\02}+1\Bigr)\,\8G_R(2,2)\nn\\
 &&\qd-\Bigl({s^2\04}+{3\02}\,s+2\Bigr)\,\8G_R(3,2)\biggr)
 +\,2\,\8G_R(1,4)-s\,\8G_R(2,4)+\Bigl({s^2\02}+s\Bigr)\,\8G_R(3,4)\nn\\
 &&-\Bigl({s^3\04}+{3\02}\,s^2+2s\Bigr)\,\8G_R(4,4)
 +\Bigl({s^4\08}+{3\02}\,s^3+{11\02}\,s^2+6s\Bigr)\,\8G_R(5,4)\nn\\
 &&+\Bigl({s^3\04}+3\,s^2+11\,s+12\Bigr)\Bigl({s^2\02}+{s\02}-6\xi_c(s^2+4s)\Bigr)
 \,\widehat{\tau_0^2\!G_R}(s;3,2)\nn\\
 &&-\Bigl({s^5\08}+{5\02}\,s^4+{35\02}\,s^3+50\,s^2+48\,s\Bigr)\,
 \widehat{\tau_0^2\!G_R}(s;5,4) \label{drhoI}\eea

For a numerical computation of $\de\r_{\widehat{I}}(s)$ we need the functions
$\8G_R(p,l)$ ((\ref{G1}) and (\ref{Ftil})) and $\widehat{G_R}(s;p,l)$ (\ref{LFb}).
According to the recurrence properties of the $Q_n$ (\ref{Qn}) they can be
easily approximated by a direct numerical summation of their corresponding
series up to the $N\,$-th term (provided that they are not alternating).
Since the terms of the series behave like $n^{-2}$ for large $n$, one
encounters an error of the order of $N^{-1}$ due to the truncation (in case of
$\widehat{G_R}(s;p,l)\,$, $N>s$ should apply). In order that this really comes
true, one has to work with a great numerical precision. For an example consider
the most difficult case $\8G_R(1,4)$:
\[ \8G_R(1,4)\;\sim\;\sum_n\Bigl({\cal O}(n^3)-{\cal O}(n^3)\Bigr)\;
 \stackrel{!}{\sim}\;\sum_n{\cal O}(n^{-2}) \]
Two terms of the order of $n^3$ calculated independently have to cancel with
an accuracy of $n^{-2}$ in the large-$n$ terms of (\ref{Ftil}). Hence a relative
accuracy of $N^{-5}$ is required for the calculation of these terms, if the
summation of the series up to $n=N$ is supposed to be sensible. Choosing
$N=10000$, a relative precision of $10^{-20}$ is required!\\
For that reason we used REAL*16 variables (29 significant digits). The
precision approximation for the gamma-function of Lanczos \cite{Lanczos} has
been extended to a relative accuracy $<10^{-25}$ in the whole complex plane
and was used for calculating the $Q_0\,$.

\subsection{Stability}
The Laplace transform of equation (\ref{Ein4}) reads ($M_{Pl}=G_N^{-1/2}$)
\beq 6\,{s\widehat{I}(s)-I(0)\0H_0}\;=\;{H_0^2\0\pi M_{Pl}^2}
 \biggl(\de\r_{\widehat{I}}
 \Bigl({s\0H_0}\Bigr)\,\widehat{I}(s)+\sum_{n=0}^2f_n\Bigl({s\0H_0}\Bigr)\,
 {I^{(n)}(0)\0H_0^{n+1}}+{1\0H_0}\,g\Bigl({s\0H_0}\Bigr)\biggr)\label{Ein5}\eeq
and is solved by
\beq \widehat{I}(s)\;=\;{6\,{I(0)\0H_0}+{H_0^2\0\pi M_{Pl}^2}
 \Bigl(\sum_{n=0}^2f_n({s\0H_0})\,
 {I^{(n)}(0)\0H_0^{n+1}}+{1\0H_0}\,g({s\0H_0})\Bigr)\06\,{s\0H_0}
 -{H_0^2\0\pi M_{Pl}^2}\,\de\r_{\widehat{I}}({s\0H_0})}\;. \label{Ihat}\eeq
Inverting the Laplace transform we get the exact general solution $I(t)$ of
the linearized ``backreaction problem'':
\beq I(t)\;=\;\int\limits_{-i\infty+\a}^{+i\infty+\a}\!{ds\02\pi i}\,
 \widehat{I}(s)\,e^{st} \label{Lainv}\eeq
The contour of integration runs on the right of all poles of the integrand
($\a>\ldots$). The numerator of $\widehat{I}$ is of the order of $s^2(1+\ln s)$
for large $|s|$ (on the negative real axis between its poles) and the
denominator is of the order of $s^3(1+\ln s)$ (see below). Therefore one can
close the contour of integration for $t>0$ by a sequence of semicircles in the
left half complex plane extended by $\a$ to the right with their centres
being located at the origin (compare with the reflected image of figure
\ref{Int}). Their radii have to be choosen in such a
way that they do not come too close to the poles of the numerator of
$\widehat{I}$ on the negative real axis. Then the contributions of these
semicircles to the integral vanish in the limit of infinite radii.
Hence the integral (\ref{Lainv}) is equal to the sum over the residues of all
poles of its integrand.\\
We are not going to perform the back-transformation explicitly, but
investigate the question of the existence of any instabilities. Due to the
factor $e^{st}$ growing terms in $I(t)$ for large $t$ are only possible
through poles at $s=s_0$ with $\re\,s_0\geq0$. By the functions
$\widehat{G_R}(s;p,l)$ the numerator of $\widehat{I}$ has infinitely many
poles (see (\ref{LFb})), but they are all located in the left half complex
plane. Thus only the zeros of the denominator of $\widehat{I}$ are of interest
to us.\\
In the following we are seeking solutions of the equation
\beq 6\,{s\0H_0}\;=\;{H_0^2\0\pi M_{Pl}^2}\,\de\r_{\widehat{I}}\Bigl({s\0H_0}
 \Bigr)\label{Nenn}\eeq
with $\re\,s\geq0\,$. Note that the factor $H_0^2/M_{Pl}^2$ has to be small
compared to 1. This is a necessary condition for the applicability of the
semiclassical theory. In the new inflationary universe
scenario for example we have $H_0\sim10^{11}\,$GeV ($M_{Pl}=10^{19}\,$GeV,
$H({\rm today})=10^{-42}\,$GeV).\\
First we will consider two hypothetical situations where
instabilities could occur:\\
$\underline{{\rm(i)}\qd\de\r_{\widehat{I}}(s)\stackrel{s\to0}{\to}\be>0}\;$
In this case a solution of (\ref{Nenn}) would be
\[ {s\0H_0}\;\simeq\;{\be\06\pi}\,{H_0^2\0M_{Pl}^2} \]
corresponding to an instability on a large time scale $s^{-1}$.
However the numerical calculations show that $\be$ is always of the order of
$N^{-1}$, which is the error due to the truncation of the hypergeometric
series at the $N\,$-th term. This observation leads to the conjecture
$\be=0$, which can be proven by a simple consideration: The constant terms
$\sim s^0$ in $\de\r_{\widehat{I}}(s)$ are coming from contributions to
$\de\r\,$ proportional to $I(t)$ which are present even for a constant $I(t)$.
A constant $I$ in the scale factor (\ref{Skal}) of the metric (\ref{Met}) can be
removed by the coordinate transform $t'=t$, $\1x'=(1+I)^{1/2}\1x$, for which
the 00-component of the energy momentum tensor (\ref{EIT5}) behaves like a scalar and
remains unchanged! Hence $s=0$ must be a solution of (\ref{Nenn}), but it
is a pure gauge mode (coordinate transform).\\
With $\de\r_{\widehat{I}}(0)=0$ we have indirectly proven the following two,
on this level remarkable identities (using (\ref{drhoI})):
\[ 2\,\8G_R(1,2)\;=\;{57\024}-{\nu^2\02\sin^2\pi\nu}
 -{\8m_0^2\0H_0^2}\Bigl({5\02}+3\g-3\ln2+{3\02}\,\psi({\ts{3\02}\!+\!\nu})
 +{3\02}\,\psi({\ts{3\02}\!-\!\nu})\Bigr)\]
\bean
 2\,\8G_R(1,4)\!&=&\!{131\016}+{49\016}\nu^2+
 {\nu^4\!-\!97\nu^2\!/16\0\sin^2\pi\nu}\\
 &&\!-\,{15\8m_0^2\04H_0^2}\biggl({3\02}+\Bigl({\8m_0^2\0H_0^2}+2\Bigr)
 \Bigl({47\060}+\g-\ln2+\2\,\psi({\ts{3\02}\!+\!\nu})
 +\2\,\psi({\ts{3\02}\!-\!\nu})\Bigr)\biggr)\eean
It is also possible to prove these identities on the level of equation
(\ref{Fk1}), but that will be skipped here.\\
The numerical result $\be=0$ can be considered a non-trivial check for the
numerical as well as analytical calculation.\\
$\underline{{\rm(ii)}\qd\de\r_{\widehat{I}}(s)\stackrel{s\gg1}{\simeq}\be s^4}
\;$ This behaviour would lead to the solution
\[ s\;\simeq\;\Bigl({6\pi\0\be}\Bigr)^{1/3}\Bigl({H_0\0M_{Pl}}\Bigr)^{1/3}
 M_{Pl} \]
and thereby to an instability on a short time scale $s^{-1}$ (compared with
$H_0^{-1}$) that still lies in the semiclassical region. However, again the
numerical investigation first showed that the $s^4\,$-terms in
$\de\r_{\widehat{I}}$ cancel each other:
\[ \begin{array}{l@{}l} {1\08}\Bigl(&
 (1\!-\!12\xi_c)(\8G_R(1,0)-s\widehat{G_R}(s;1,0))-(\8G_R(3,2)
 -s\widehat{G_R}(s;3,2))\\
 &-(1\!-\!12\xi_c)(\8G_R(3,2)-s\widehat{\tau_0^2\!G_R}(s;3,2))+\8G_R(5,4)
 -s\widehat{\tau_0^2\!G_R}(s;5,4)\Bigr)\\
 \lefteqn{\sim\,{\cal O}\Bigl(s^{-1}(1+\ln s)\Bigr)}\end{array} \]
The reason is that fourth derivatives $I^{\rm(iv)}(t)$ do emerge only in the
convolution integrals (\ref{Fki}) from the singular behaviour of the kernels
$J^{(4)}_l\,$. Using (\ref{Jas}) one sees that in fact in (\ref{EIT6}) the
corresponding terms from $\4\6_t^2\int\!\8d\8k\,\de(2\,\re A)^{-1}$ and from
$\int\!\8d\8k\,\de(a^{-2}k^2/2\,\re A)$ cancel.\\
More easily it follows from the energy momentum conservation
$\9{\de\r}+3H_0(\de\r+\de p)=0\,$, that $\de\r$ has to contain one time
derivative less than $\de p$ ($\9{\r}_0=0=\r_0+p_0$, see (\ref{EIT5})).\\
Summarizing we did not find an instability but again a non-trivial check on the
numerical as well as analytical calculation.
\begin{figure}[bp]
 \vspace{10.7cm}
 \includegraphics{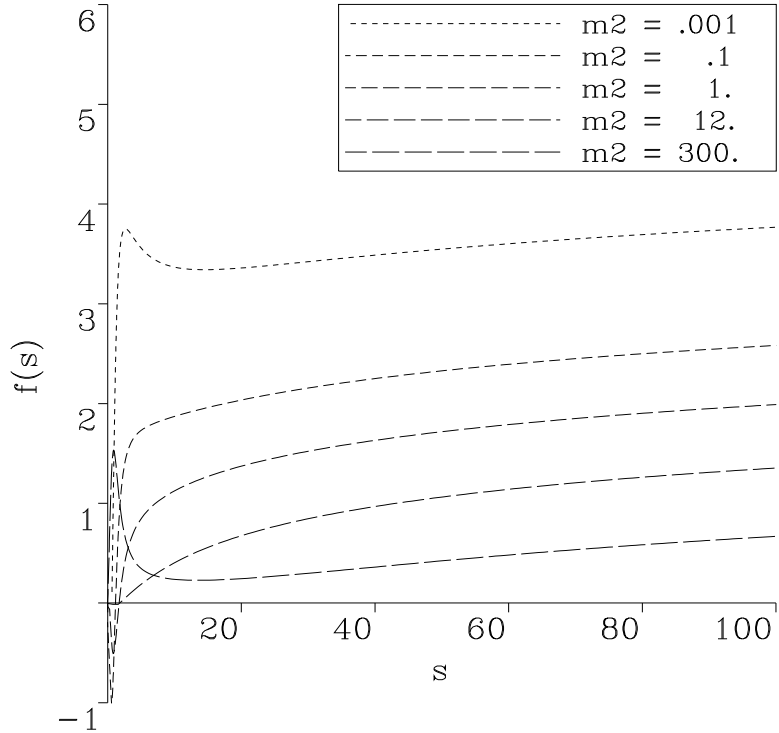}
 \caption{The function $f(s):=\de\r_{\widehat{I}}(s)/(s^3+1)$ for $\xi=0$ and
   certain values of $m2:=m^2/H_0^2$. For $m2=300$ a peak appears on the left
   which is due to numerical inaccuracies. Contrary to the rest of the
   curve it depends strongly on the numerical precision $N^{-1}$.}
 \label{Plott1}
\end{figure}

\begin{figure}[bp]
 \vspace{10.7cm}
 \includegraphics{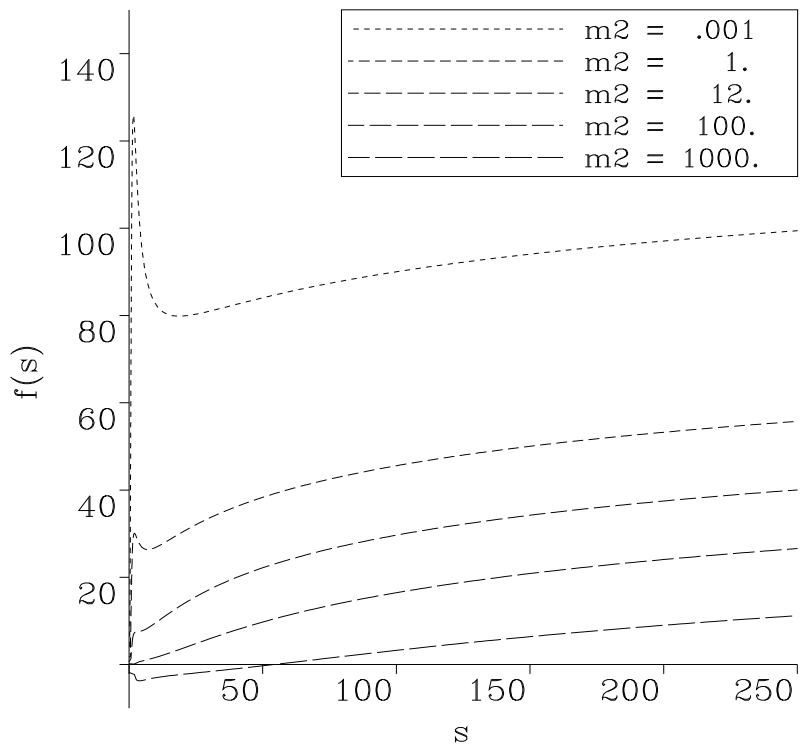}
 \caption{The function $f(s):=\de\r_{\widehat{I}}(s)/(s^3+1)$ for $\xi=1$ and
          certain values of $m2:=m^2/H_0^2$}
 \label{Plott2}
\end{figure}

Now it's time to stop making hypothesises and to look at the real behaviour of
$\de\r_{\widehat{I}}(s)$. The function $\de\r_{\widehat{I}}(s)/(s^3+1)$ has
been plotted in the figures \ref{Plott1} and \ref{Plott2} for two values of
$\xi$ and different values of $m^2$ for real positive $s$. The curves are
growing only logarithmically for large $s$.

It has been verified and can also be seen in (\ref{drhoI}) that the slope of
$\de\r_{\widehat{I}}$ near $s=0$ does not reach an excessive large numerical
value. Since $H_0^2/M_{Pl}^2$ is small, a large $s$ only can therefore solve
equation (\ref{Nenn}). According to our considerations so far and to the
numerical investigations $\de\r_{\widehat{I}}(s)$ behaves like
$\a s^3(\ln s+\be)$ for large $s$, so that (\ref{Nenn}) has the following
solution:
\beq s\;=\;\Bigl({6\pi\0\a'}\Bigr)^{1/2}\,M_{Pl} \label{Pls}\eeq
$\a'$ depends on $\a$ and $\be$ without receiving overwhelming large numerical
values. Hence this solution would lead to an instability on the Planck time
scale, which is apart from the region of validity of the semiclassical theory.
Within our semiclassical treatment we are not able to conclude for an
instability on the Planck scale. The investigation of this region remains a
subject for a future quantum theory of gravity.\\
For a constant ratio $m^2/H_0^2$ the solution (\ref{Pls}) does not depend on
$H_0$ (apart from logarithmic terms), so that it would be present even for very
small $H_0\,$. Merely the existence of our present universe seems therefore to
exclude an instability of this kind within a complete, applying theory.
Probably it is just an artefact of the semiclassical treatment, as was argued
for example in reference \cite{IsaRog}.\\
For the practical use of our general semiclassical solution it is always
possible to avoid this Planck mode by requiring the numerator
of (\ref{Ihat}) to have a zero at the same value (\ref{Pls}), too. This condition
is one constraint in the space of initial data which were discussed in section
\ref{Laplace}. The general solution then consists of an infinite series
of exponentially damped modes due to the poles from (\ref{Ihat}) with
$\re\,s<0$ and a constant mode from the $s=0$ pole. The last one is the only
one to survive for late times, but it corresponds just to a spatial rescaling
of the underlying de~Sitter spacetime with no influence on physical
observables (like its curvature for example).
\expandafter\ifx\csname ok\endcsname\relax
   \end{document}\fi
\expandafter\ifx\csname qd\endcsname\relax
   \documentstyle[12pt,twoside]{article}\let\qd=\quad \let\qqd=\qquad \def\qqqd{\qquad\qquad}

\let\a=\alpha \let\be=\beta \let\g=\gamma \let\de=\delta
\let\e=\varepsilon \let\z=\zeta \let\h=\eta \let\th=\theta
\let\dh=\vartheta \let\k=\kappa \let\la=\lambda
\let\r=\rho \let\s=\sigma
\let\om=\omega
\let\ph=\varphi
\let\Om=\Omega \let\Si=\Sigma \let\Th=\Theta
\let\La=\Lambda \let\G=\Gamma \let\D=\Delta

\def\0{\over } \def\1{\vec } \def\2{{1\over2}} \def\4{{1\over4}}
\def\5{\bar } \def\6{\partial } \def\7#1{{#1}\llap{/}}
\def\8{\tilde } \def\9{\dot }

\def\ul#1{$\underline{\hbox{#1}}$} \def\llp{\hbox to 0pt{\hss /\hskip1.5pt}}
\def\so{\supset\hbox to 0pt{\hss $\displaystyle -$\hskip1pt}}

\def\ts{\textstyle } \def\ds{\displaystyle }
\def\<{\langle } \def\>{\rangle } \def\lb{\left} \def\rb{\right}
\let\lra=\leftrightarrow \let\LRA=\Leftrightarrow
\let\then=\Rightarrow \def\triang{{\scriptstyle \Delta}}
\def\trian{{\scriptscriptstyle \Delta}}
\def\re{{\rm Re}} \def\im{{\rm Im}} \def\ord{{\cal O}}

\let\ap=\approx \let\eq=\equiv \let\ex=\times \let\hc=\dagger
      \let\ti=\tilde \let\bl=\biggl \let\br=\biggr
\let\bi=\choose \let\at=\atop \let\mat=\pmatrix

\let\nn=\nonumber
\def\bea{\begin{eqnarray}} \def\eea{\end{eqnarray}}
\def\bean{\begin{eqnarray*}} \def\eean{\end{eqnarray*}}
\def\beq{\begin{equation}} \def\eeq{\end{equation}}
\def\barr{\begin{array}} \def\earr{\end{array}}

\newcommand{\deflabel}[1]{#1\hfil}%
\newenvironment{deflist}[1]%
{\begin{list}{}%
{\settowidth{\labelwidth}{#1}%
\setlength{\leftmargin}{\labelwidth}\addtolength{\leftmargin}{\labelsep}%
\renewcommand{\makelabel}{\deflabel}}}%
{\end{list}}%
\fi
\expandafter\ifx\csname ok\endcsname\relax
   \def\re{{\rm Re}} \def\im{{\rm Im}}
   {\catcode`@=11 \@addtoreset{equation}{section}\@addtoreset{figure}{section}}
   \def\theequation{\thesection .\arabic{equation}}
   \def\thefigure{\thesection .\arabic{figure}}
   \addtolength{\topmargin}{-48pt}
   \addtolength{\textheight}{90pt}
   \addtolength{\evensidemargin}{-31pt}
   \addtolength{\oddsidemargin}{-12pt}
   \addtolength{\textwidth}{30pt}
   \addtolength{\footskip}{21pt}
   \setlength{\parindent}{0pt}\frenchspacing
   \begin{document}
\fi
\section{Summary and conclusions}
We have linearized the semiclassical system of Schr\"odinger equation and
Einstein equations (\ref{EinSchro}) around the de~Sitter -- Bunch-Davies solution
in order to investigate the stability of this solution against small
fluctuations of the quantum state and small, spatially homogeneous and
isotropic perturbations of the k=0 FRW de~Sitter metric. This linearization
in the sense of a stability analysis is the only approximation appearing
in the present work.\\
The condition of finite energy density for the initial quantum state has been
discussed and the Schr\"odinger equation was completely solved. The expectation
value of the energy momentum tensor has been calculated as a functional of the
metric perturbation. The momentum integrations have been carried out
analytically. A procedure for the isolation of divergencies was developed.
These were removed by a renormalization through the subtraction of
De~Witt-Schwinger terms.\\
The linearized semiclassical Einstein equations became a linear
integro-differential equation for the metric perturbation. After finding out
the initial data which can be specified the general solution was obtained in
terms of its Laplace transform. This Laplace transform has been analyzed
analytically as well as numerically using the necessary high numerical
precision. The general solution contains only two potential instabilities:
a constant mode which corresponds just to a constant spatial rescaling of the
underlying de~Sitter spacetime i.e. a pure gauge, and an instability on the
Planck time scale which is outside of the scope of a semiclassical theory.\\
Thus we have shown that de~Sitter spacetime and Bunch-Davies vacuum are stable
within our semiclassical theory!\\
The same result was obtained in reference \cite{IsaRog} for the special case of
minimal coup\-ling $\xi=0$. Since there the momentum integrals are not evaluated
explicitly, complicated estimates were necessary in order to achieve this and
there is no numerical analysis.\\
Our conclusion above is in contradiction with some claims existing in the
literature. In particular in ref. \cite{Mottola} an instability on the Hubble
time scale $H_0^{-1}$ was found. They use indeed a different coordinate system
(k=+1 FRW), but the coordinate lines $t=\hbox{const.}$, on which the spatially
homogeneous initial data of the perturbation are specified, tend to coincide
within the k=0 and k=+1 FRW parametrizations for late times $t\to\infty$.
Therefore the answer to the stability issue should be the same.
Some criticism on appendix A of ref. \cite{Mottola} is already contained in
ref. \cite{IsaRog}. Furthermore the diverging perturbation $\s(\eta)$ of the
conformal factor found in ref. \cite[section 5]{Mottola} is very near to a
pure gauge mode. It can be transformed into our $I(t)$ remaining
small for all times and contains no physical divergence.\\
Finally a few possibilities for a continuation of this work should be
mentioned: Fermi fields could be included, but presumably they wouldn't alter
the stability argumentation. The investigation of the minimally coupled
massless case would be interesting, because the renormalized two point function
in de~Sitter spacetime (see refs. \cite{Allen, FoVi, AlFo})
as well as our result (\ref{drhoI}) are containing a
logarithmic infrared divergence (the linear infrared divergencies of the
$\psi\,$- and $G\,$-functions cancel). At last one could try to access the
region of quantum gravity via the Wheeler-De~Witt equation.\\[10mm]
I would like to thank W. Buchm\"uller for many helpful discussions in the
course of this work and D. Litim for many valuable comments on the
manuscript.

\expandafter\ifx\csname ok\endcsname\relax
   \end{document}\fi
\clearpage
{\catcode`@=11 \@addtoreset{equation}{section}\@addtoreset{figure}{section}}
\def\theequation{\thesection .\arabic{equation}}
\def\thefigure{\thesection .\arabic{figure}}
\expandafter\ifx\csname qd\endcsname\relax
   \documentstyle[12pt,twoside]{article}\let\qd=\quad \let\qqd=\qquad \def\qqqd{\qquad\qquad}

\let\a=\alpha \let\be=\beta \let\g=\gamma \let\de=\delta
\let\e=\varepsilon \let\z=\zeta \let\h=\eta \let\th=\theta
\let\dh=\vartheta \let\k=\kappa \let\la=\lambda
\let\r=\rho \let\s=\sigma
\let\om=\omega
\let\ph=\varphi
\let\Om=\Omega \let\Si=\Sigma \let\Th=\Theta
\let\La=\Lambda \let\G=\Gamma \let\D=\Delta

\def\0{\over } \def\1{\vec } \def\2{{1\over2}} \def\4{{1\over4}}
\def\5{\bar } \def\6{\partial } \def\7#1{{#1}\llap{/}}
\def\8{\tilde } \def\9{\dot }

\def\ul#1{$\underline{\hbox{#1}}$} \def\llp{\hbox to 0pt{\hss /\hskip1.5pt}}
\def\so{\supset\hbox to 0pt{\hss $\displaystyle -$\hskip1pt}}

\def\ts{\textstyle } \def\ds{\displaystyle }
\def\<{\langle } \def\>{\rangle } \def\lb{\left} \def\rb{\right}
\let\lra=\leftrightarrow \let\LRA=\Leftrightarrow
\let\then=\Rightarrow \def\triang{{\scriptstyle \Delta}}
\def\trian{{\scriptscriptstyle \Delta}}
\def\re{{\rm Re}} \def\im{{\rm Im}} \def\ord{{\cal O}}

\let\ap=\approx \let\eq=\equiv \let\ex=\times \let\hc=\dagger
      \let\ti=\tilde \let\bl=\biggl \let\br=\biggr
\let\bi=\choose \let\at=\atop \let\mat=\pmatrix

\let\nn=\nonumber
\def\bea{\begin{eqnarray}} \def\eea{\end{eqnarray}}
\def\bean{\begin{eqnarray*}} \def\eean{\end{eqnarray*}}
\def\beq{\begin{equation}} \def\eeq{\end{equation}}
\def\barr{\begin{array}} \def\earr{\end{array}}

\newcommand{\deflabel}[1]{#1\hfil}%
\newenvironment{deflist}[1]%
{\begin{list}{}%
{\settowidth{\labelwidth}{#1}%
\setlength{\leftmargin}{\labelwidth}\addtolength{\leftmargin}{\labelsep}%
\renewcommand{\makelabel}{\deflabel}}}%
{\end{list}}%
\fi
\expandafter\ifx\csname ok\endcsname\relax
   \def\re{{\rm Re}} \def\im{{\rm Im}}
   {\catcode`@=11 \@addtoreset{equation}{section}\@addtoreset{figure}{section}}
   \def\theequation{\thesection.\arabic{equation}}
   \def\thefigure{\thesection.\arabic{figure}}
   \addtolength{\topmargin}{-48pt}
   \addtolength{\textheight}{90pt}
   \addtolength{\evensidemargin}{-31pt}
   \addtolength{\oddsidemargin}{-12pt}
   \addtolength{\textwidth}{30pt}
   \addtolength{\footskip}{21pt}
   \setlength{\parindent}{0pt}\frenchspacing
   \def\vFd{{}_4\5F_3}
   \begin{document}
\fi
\setcounter{section}{0} \def\thesection{\Alph{section}}

\section{Integrals of products of Hankel functions}\label{Hankel}
\begin{figure}[bp]
 \includegraphics{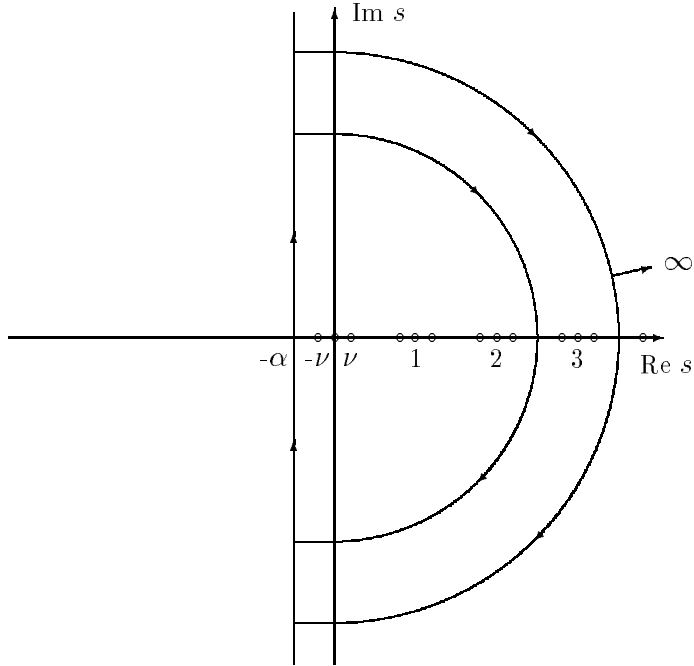}
 \vspace{9.2cm}
 \caption{The contour of integration in the complex $s\,$-plane is being
   closed by a sequence of semicircles, whose radii tend to infinity. The
   poles of the integrand in (\protect\ref{H12}) are marked on the real axis.}
 \label{Int}
 \vspace{5mm}
\end{figure}

The definite integral of a product of two Hankel functions can be evaluated
using the Weber-Schafheitlin integral \cite{Wat}.
One obtains:
\bea \int\limits_0^{\infty}\!dx\,x^{\la}H^{(1)}_{\nu}(x)\,H^{(2)}_{\nu}(x)
 \!&=&\!\!
 \pi^{-{5\02}}\,\G\Bigl(-{\la\02}\Bigr)\,\G\Bigl({1\!+\!\la\02}\Bigr)\cos\nu\pi\,
 \G\Bigl({1\!+\!\la\02}+\nu\Bigr)\,\G\Bigl({1\!+\!\la\02}-\nu\Bigr)\nn\\
 \label{WebSchaf3}\\
 \int\limits_0^{\infty}\!dx\,x^{\la}{H^{(1)}_{\nu}}^2\!(x)\!&=&\!\!
 -i\,\pi^{-{3\02}}\,e^{-i\pi(\nu-{\la\02})}\,{\G({1+\la\02})\0\G(1\!+\!{\la\02})}\,
 \G\Bigl({1\!+\!\la\02}+\nu\Bigr)\,\G\Bigl({1\!+\!\la\02}-\nu\Bigr)\nn\\
 \label{WebSchaf1}\\
 \int\limits_0^{\infty}\!dx\,x^{\la}{H^{(2)}_{\nu}}^2\!(x)\!&=&\!\!
 i\,\pi^{-{3\02}}\,e^{i\pi(\nu-{\la\02})}\,{\G({1+\la\02})\0\G(1\!+\!{\la\02})}\,
 \G\Bigl({1\!+\!\la\02}+\nu\Bigr)\,\G\Bigl({1\!+\!\la\02}-\nu\Bigr)\nn\\
 \label{WebSchaf2}\eea
Some functional relations for the gamma-function (duplication and supplement)
have been applied. The integrals (\ref{WebSchaf3})--(\ref{WebSchaf2}) are absolutely convergent within
the region $2\,|\re\,\nu|-1<\re\,\la<0$. The left inequality stems from the
behaviour of the integrand for small $x$ and
secures infrared convergence. From the asymptotic behaviour of the Hankel functions the right
inequality follows, which stands for ultraviolet convergence. The integrals
(\ref{WebSchaf1}) and (\ref{WebSchaf2}) are convergent even for $\re\,\la<1$
due to the oscillating behaviour of the Hankel functions.\\
In section \ref{Fast} we need the following integral of a product of four
Hankel functions:
\beq {\cal I}\;:=\;\int\limits_0^{\infty}\!dx\,x^{\la-1}{H^{(1)}_{\nu}}^2\!
 (ax)\,{H^{(2)}_{\nu}}^2\!(x)\;,\qd 0<a\leq1 \label{Int40}\eeq
Such an integral of four Bessel functions was not found in the mathematical
standard literature. Hence it will be evaluated explicitly now.
The integrand behaves like $\,x^{\la-1-4|\re\,\nu|}\,$ for $x\to0$ and like
$\,x^{\la-3}\,$ for $x\to\infty$, so that the integral (\ref{Int40}) is
absolutely convergent in the region $4\,|\re\,\nu|<\re\,\la<2$. If $a\neq1$
it is convergent even for $\re\,\la<3\,$ due to the oscillating behaviour of the
integrant. We suppose that we are within this region since then all integrals
and series appearing in the following will be convergent, too.\\
For the square of $H^{(1)}_{\nu}(x)$ we use an integral representation:
\beq {H^{(1)}_{\nu}}^2\!(x)\,=\,{-2\0\pi^{3\02}}\,e^{-i\pi\nu}
 \int\limits_{-i\infty-\a}^{+i\infty-\a}\!{ds\02\pi i}\,x^{2s}e^{-i\pi s}
 {\G(-s)\,\G(-s-\nu)\,\G(-s+\nu)\0\G(\2-s)} \label{H12}\eeq
$\forall\,x\in\hbox to 1.7pt{$I$\hss}R^{+}\,$ and $\;|\re\,\nu|<\a<{3\04}\,$.
Integrals of this type are called Mellin-Barnes integrals. In order to prove
(\ref{H12}) we close the contour of integration by a sequence of semicircles
in the right half complex plane extended by $\a$ to the left with their
centres in the origin according to figure \ref{Int}.
The radii of these semicircles have to be choosen in such a way that they do
not come too close to the poles of the three gamma-functions in the numerator.
Using functional relations and Stirling's asymptotic expansion for the gamma-function
one shows similar to \cite[\S14.5]{WhittWat} that the contributions of the
semicircles to the integral are vanishing in the limit of an infinite radius.
Thus the theorem of residues can be used to evaluate the integral (\ref{H12}).
The gamma-function $\G(s)$ has simple poles at all negative integers
$s=-n$, $n\in\hbox to 1.5pt{$I$\hss}N_0$ with the residues
$(-)^n/n!\,$. The integral (\ref{H12}) becomes the sum of the residues of all
poles of the three gamma-functions in the numerator of the integrand:
\bean {H^{(1)}_{\nu}}^2\!(x)\!&=&\!{-2\0\pi^{3\02}}\,e^{-i\pi\nu}
 \sum_{n=0}^{\infty}{(-)^n\0n!}\biggl(x^{2n}e^{-i\pi n}\,
 {\G(-n+\nu)\,\G(-n-\nu)\0\G(\2-n)}\\
 &&\qqqd+\;x^{2(n-\nu)}e^{-i\pi(n-\nu)}\,{\G(-n+\nu)\,\G(-n+2\nu)\0\G(\2-n+\nu)
 }\nn\\ &&\qqqd+\;
 x^{2(n+\nu)}e^{-i\pi(n+\nu)}\,{\G(-n-\nu)\,\G(-n-2\nu)\0\G(\2-n-\nu)}\biggr)
\eean
Via functional relations for the gamma-function and the
power series representation \cite[\S5.41]{Wat} of the product of two Bessel
functions $J_{\nu}$ the proof of (\ref{H12}) can be completed:
\bean {H^{(1)}_{\nu}}^2\!(x)\!
 &=&\!{-1\0\sin^2\nu\pi}\biggl(-2\,e^{-i\pi\nu}\sum_{n=0}^{\infty}
 {(-)^n\;\G(1+2n)\;(x/2)^{2n}\0n!\,\G(1-\nu+n)\,\G(1+\nu+n)\,\G(1+n)}\\
 &&\qqd+\;\sum_{n=0}^{\infty}{(-)^n\;\G(1-2\nu+2n)\;(x/2)^{2n-2\nu}\0
 n!\,\G(1-\nu+n)\,\G(1-2\nu+n)\,\G(1-\nu+n)}\\
 &&\qqd+\;e^{-2i\pi\nu}\sum_{n=0}^{\infty}{(-)^n\;\G(1+2\nu+2n)\;(x/2)^{2n+2\nu}\0
 n!\,\G(1+\nu+n)\,\G(1+2\nu+n)\,\G(1+\nu+n)}\biggr)\\
 &=&\!{-1\0\sin^2\nu\pi}\,\Bigl(-2\,e^{-i\pi\nu}\,J_{\nu}J_{-\nu}+J_{-\nu}^{\,2}
 +e^{-2i\pi\nu}\,J_{\nu}^{\,2}\Bigr)\qqd{\rm q.e.d.}\eean
With the integral representation (\ref{H12}) our integral (\ref{Int40}) reads
after interchanging the order of integration:
\[ {\cal I}\,=\,{-2\0\pi^{3\02}}\,e^{-i\pi\nu}
 \int\limits_{-i\infty-\a}^{+i\infty-\a}\!{ds\02\pi i}\,a^{2s}e^{-i\pi s}
 {\G(-s)\,\G(-s-\nu)\,\G(-s+\nu)\0\G(\2-s)}\int\limits_0^{\infty}\!dx\,
 x^{\la-1+2s}\,{H^{(2)}_{\nu}}^2\!(x) \]
Choosing $\a$ such that $\re\,\la-2<2\a<\re\,\la-2\,|\re\,\nu|\,$ we put
in the Weber-Schafheitlin integral (\ref{WebSchaf2}) and obtain:
\bea {\cal I}\!&=&\!{2\0\pi^3}\,(-i)^{\la}
 \int\limits_{-i\infty-\a}^{+i\infty-\a}\!{ds\02\pi i}\,(a^2e^{-2\pi i})^s\nn\\
 &&\cdot\;{\G(-s)\,\G(-s-\nu)\,\G(-s+\nu)\0\G(\2-s)}\,
 {\G({\la\02}+s)\,\G({\la\02}+s+\nu)\,\G({\la\02}+s-\nu)\0\G({\la+1\02}+s)}
 \nn\\ \label{MeierG}\eea
For $a\leq1$ we can close the contour of integration again by a sequence of
extended semicircles in the right half complex plane with their radii tending
to infinity (figure \ref{Int}). From the theorem of residues we find:
\bea {\cal I}\!&=&\!{(-i)^{\la}\0\pi^2\sin^2\nu\pi}\sum_{n=0}^{\infty}
 {\G({\la\02}+n)\0n!}\biggl(-2\,a^{2n}\,{\G({\la\02}-\nu+n)\,\G({\la\02}+\nu+n)
 \,\G(\2+n)\0\G(1+\nu+n)\,\G(1-\nu+n)\,\G({\la+1\02}+n)}\qd\nn\\
 &&\qqd+\;a^{2n-2\nu}e^{2\pi i\nu}\,{\G({\la\02}-\nu+n)\,\G({\la\02}-2\nu+n)
 \,\G(\2-\nu+n)\0\G(1-\nu+n)\,\G(1-2\nu+n)\,\G({\la+1\02}-\nu+n)}\nn\\
 &&\qqd+\;a^{2n+2\nu}e^{-2\pi i\nu}\,{\G({\la\02}+\nu+n)\,\G({\la\02}+2\nu+n)
 \,\G(\2+\nu+n)\0\G(1+\nu+n)\,\G(1+2\nu+n)\,\G({\la+1\02}+\nu+n)}\biggr)\nn\\
 &=&\!{(-i)^{\la}\0\pi^2\sin^2\nu\pi}\biggl(-2\,\vFd\Bigl({\la\02},{\la\02}
 \!-\!\nu,{\la\02}\!+\!\nu,\2;1\!+\!\nu,1\!-\!\nu,{\la\!+\!1\02};a^2\Bigr)\nn\\
 &&\qd+\;a^{-2\nu}e^{2\pi i\nu}\,\vFd\Bigl({\la\02}\!-\!\nu,{\la\02}\!-\!2\nu,
 {\la\02},\2\!-\!\nu;1\!-\!\nu,1\!-\!2\nu,{\la\!+\!1\02}\!-\!\nu;a^2\Bigr)\nn\\
 &&\qd+\;a^{2\nu}e^{-2\pi i\nu}\,\vFd\Bigl({\la\02}\!+\!\nu,{\la\02},{\la\02}
 \!+\!2\nu,\2\!+\!\nu;1\!+\!\nu,1\!+\!2\nu,{\la\!+\!1\02}\!+\!\nu;a^2\Bigr)
 \biggr)\eea
We have applied functional relations and introduced our generalized hypergeometric
function
\bea {}_p\5F_q(\a_1,\ldots,\a_p;\be_1,\ldots,\be_q;z)\!&:=&\!\sum_{n=0}^{\infty}
 {\G(\a_1+n)\,\cdots\,\G(\a_p+n)\0n!\,\G(\be_1+n)\,\cdots\,\G(\be_q+n)}\;z^n
 \label{Fdef}\\
 &=&\!{\G(\a_1)\cdots\G(\a_p)\0\G(\be_1)\cdots\G(\be_q)}\,{}_pF_q(\a_1,\ldots,
 \a_p;\be_1,\ldots,\be_q;z),\nn\eea
where ${}_pF_q$ is the usual generalized hypergeometric
function.\\
Note that (\ref{MeierG}) is an integral of the Mellin-Barnes type, too, and is
therefore a special case of the general notion of Meier's $G$-function
\cite{HTI}:
\[ {\cal I}\,=\,{2\0\pi^3}\,(-i)^{\la}\,G^{33}_{44}\!\lb(a^2e^{-2\pi i}\,
 \begin{array}{|c@{,\,}c@{,\,}c@{,\,}c}
   1\!-\!{\la\02}&1\!-\!{\la\02}\!+\!\nu&1\!-\!{\la\02}\!-\!\nu&\2\\
   0&\nu&-\nu&{1-\la\02}\end{array}\rb) \]
In section \ref{Fast} we have $\la=d+l$ and $a=e^{-t}\,$, and it turns out to
be reasonable to introduce the following function $G$:
\bea \lefteqn{G(t;p,l)\;:=\;{-e^{-td}\04\sin^2\pi\nu}\biggl(
  -2\,\vFd\Bigl(\hbox{\small$\ds
  {d\!+\!l\02}\!-\!p,{d\!+\!l\02}\!-\!\nu,{d\!+\!l\02}\!+\!\nu,\2;
  1\!+\!\nu,1\!-\!\nu,{d\!+\!l\!+\!1\02};e^{-2t}$}\Bigr)}\nn\\
  &&\!\!+\,e^{2\nu t}\,e^{2\pi i\nu}\vFd
  \Bigl(\hbox{\small$\ds{d\!+\!l\02}\!-\!\nu\!-\!p,
  {d\!+\!l\02}\!-\!2\nu,{d\!+\!l\02},\2\!-\!\nu;1\!-\!\nu,1\!-\!2\nu,
  {d\!+\!l\!+\!1\02}\!-\!\nu;e^{-2t}$}\Bigr)\nn\\
  &&\!\!+\,e^{-2\nu t}\,e^{-2\pi i\nu}\vFd
  \Bigl(\hbox{\small$\ds{d\!+\!l\02}\!+\!\nu\!-\!p,
  {d\!+\!l\02},{d\!+\!l\02}\!+\!2\nu,\2\!+\!\nu;1\!+\!\nu,1\!+\!2\nu,
  {d\!+\!l\!+\!1\02}\!+\!\nu;e^{-2t}$}\Bigr)\biggr)\nn\\
  \label{Gdef}\eea
Then our integral (\ref{Int40}) reads just
\beq e^{-td}\,{\cal I}\;=\;-\,{4\0\pi^2}\,(-i)^{d+l}\,G(t;0,l)\,.\label{AJ4l}\eeq
With the auxiliary variable $p$ it is possible to express the time derivative
of a $G$-function in terms of $G$-functions:
\beq \biggl(\2\,\6_t+p+1-{l\02}\biggr)\,G(t;p\!+\!1,l)\,=\,-\,G(t;p,l)\eeq
This relation can simply be proven by substituting the definitions (\ref{Gdef}) and
(\ref{Fdef}).

\expandafter\ifx\csname ok\endcsname\relax
   \end{document}\fi

\expandafter\ifx\csname qd\endcsname\relax
   \documentstyle[12pt,twoside]{article}\let\qd=\quad \let\qqd=\qquad \def\qqqd{\qquad\qquad}

\let\a=\alpha \let\be=\beta \let\g=\gamma \let\de=\delta
\let\e=\varepsilon \let\z=\zeta \let\h=\eta \let\th=\theta
\let\dh=\vartheta \let\k=\kappa \let\la=\lambda
\let\r=\rho \let\s=\sigma
\let\om=\omega
\let\ph=\varphi
\let\Om=\Omega \let\Si=\Sigma \let\Th=\Theta
\let\La=\Lambda \let\G=\Gamma \let\D=\Delta

\def\0{\over } \def\1{\vec } \def\2{{1\over2}} \def\4{{1\over4}}
\def\5{\bar } \def\6{\partial } \def\7#1{{#1}\llap{/}}
\def\8{\tilde } \def\9{\dot }

\def\ul#1{$\underline{\hbox{#1}}$} \def\llp{\hbox to 0pt{\hss /\hskip1.5pt}}
\def\so{\supset\hbox to 0pt{\hss $\displaystyle -$\hskip1pt}}

\def\ts{\textstyle } \def\ds{\displaystyle }
\def\<{\langle } \def\>{\rangle } \def\lb{\left} \def\rb{\right}
\let\lra=\leftrightarrow \let\LRA=\Leftrightarrow
\let\then=\Rightarrow \def\triang{{\scriptstyle \Delta}}
\def\trian{{\scriptscriptstyle \Delta}}
\def\re{{\rm Re}} \def\im{{\rm Im}} \def\ord{{\cal O}}

\let\ap=\approx \let\eq=\equiv \let\ex=\times \let\hc=\dagger
      \let\ti=\tilde \let\bl=\biggl \let\br=\biggr
\let\bi=\choose \let\at=\atop \let\mat=\pmatrix

\let\nn=\nonumber
\def\bea{\begin{eqnarray}} \def\eea{\end{eqnarray}}
\def\bean{\begin{eqnarray*}} \def\eean{\end{eqnarray*}}
\def\beq{\begin{equation}} \def\eeq{\end{equation}}
\def\barr{\begin{array}} \def\earr{\end{array}}

\newcommand{\deflabel}[1]{#1\hfil}%
\newenvironment{deflist}[1]%
{\begin{list}{}%
{\settowidth{\labelwidth}{#1}%
\setlength{\leftmargin}{\labelwidth}\addtolength{\leftmargin}{\labelsep}%
\renewcommand{\makelabel}{\deflabel}}}%
{\end{list}}%
\fi
\expandafter\ifx\csname ok\endcsname\relax
   {\catcode`@=11 \@addtoreset{equation}{section}\@addtoreset{figure}{section}}
   \def\theequation{\thesection.\arabic{equation}}
   \def\thefigure{\thesection.\arabic{figure}}
   \addtolength{\topmargin}{-48pt}
   \addtolength{\textheight}{90pt}
   \addtolength{\evensidemargin}{-31pt}
   \addtolength{\oddsidemargin}{-12pt}
   \addtolength{\textwidth}{30pt}
   \addtolength{\footskip}{21pt}
   \setlength{\parindent}{0pt}\frenchspacing
   \begin{document}
\fi

\def\dH{H\raise10.1pt\hbox to 0pt{\hss.\hskip-1.44pt.\hskip-1.44pt.\hskip0.5pt}}
\section[Geometrical tensors for the k=0 FRW-metric]%
{Geometrical tensors for the k=0\protect\\ FRW-metric}\label{geoT}
The k=0 FRW-metric
\beq g_{00}\;=\;1\,,\qqd g_{0i}\;=\;0\,,\qqd g_{ij}\;=\;-a^2(t)\,\de_{ij} \eeq
leads to the Christoffel symbols
\beq \G_{ij}^0\;=\;H\,a^2\de_{ij}\,,\qqd \G_{j0}^i\;=\;\G_{0j}^i\;=\;
 H\,\de^i_j\,,\qqd 0\;\mbox{otherwise}\,,\qd H=\9a/a\,. \eeq
From them we obtain the Riemann tensor
\bea {R^{\mu}}_{\nu\r\s}\!&:=&\!\6_{\r}\G_{\nu\s}^{\mu}-\6_{\s}\G_{\nu\r}^{\mu}
  +\G_{\r\la}^{\mu}\G_{\s\nu}^{\la}-\G_{\s\la}^{\mu}\G_{\r\nu}^{\la}\;:\nn\\
  {R^{0}}_{i0j}\!&=&\!-{R^{0}}_{ij0}\;=\;a^2(\9H+H^2)\,\de_{ij}\,,\qd
  {R^{i}}_{00j}\;=\;-{R^{i}}_{0j0}\;=\;(\9H+H^2)\,\de^i_j\,,\nn\\
  {R^{i}}_{kjl}\!&=&\!a^2H^2(\de^i_j\de_{kl}-\de^i_l\de_{kj})\,,\qqd
   0\;\mbox{otherwise}\,, \eea
as well as the Ricci tensor and curvature scalar
\bea R_{\mu\nu}\!&:=&\!{R^{\la}}_{\mu\la\nu}\;:\qd R_{00}\;=\;-d\,(\9H+H^2)\,,
  \qd R_{ij}\;=\;(\9H+d\,H^2)\,a^2\de_{ij}\,,\qd R_{0i}\;=\;0\nn\\
  R\!&:=&\!g^{\mu\nu}R_{\mu\nu}\;=\;-d\,(2\9H+(d\!+\!1)\,H^2)\,. \eea
The $H$-tensors read:
\bea H_{\mu\nu}&:=&{1\0\sqrt{|g|}}\,{\de\0\de g^{\mu\nu}}\int d^{d\!+\!1}\!x\,
   \sqrt{|g|}\,R^{\a\be\r\s}R_{\a\be\r\s}\nn\\
   &=&-\2g_{\mu\nu}R^{\a\be\r\s}R_{\a\be\r\s}+2R_{\mu\a\be\r}{R_{\nu}}^{\a\be\r}
   +4\Box R_{\mu\nu}-2R_{;\mu\nu}-4R_{\mu\a}{R^{\a}}_{\nu}\nn\\
   & &+\;4R^{\a\be}R_{\a\mu\be\nu}\nn\\
^{(1)}\!H_{\mu\nu}&:=&{1\0\sqrt{|g|}}\,{\de\0\de g^{\mu\nu}}\int d^{d\!+\!1}\!x\,
   \sqrt{|g|}\,R^2\nn\\
   &=&-\2g_{\mu\nu}R^2+2R\,R_{\mu\nu}-2R_{;\mu\nu}+2g_{\mu\nu}\Box R\nn\\
^{(2)}\!H_{\mu\nu}&:=&{1\0\sqrt{|g|}}\,{\de\0\de g^{\mu\nu}}\int d^{d\!+\!1}\!x\,
   \sqrt{|g|}\,R^{\a\be}R_{\a\be}\nn\\
   &=&-\2g_{\mu\nu}R^{\a\be}R_{\a\be}+2{R_{\mu}}^{\a}R_{\a\nu}
-2{{R_{\mu}}^{\a}}_{;\nu\a}+\Box R_{\mu\nu}+\2g_{\mu\nu}\Box R \label{Hmunu}\eea
Their non-zero components are:
\bea H_{00}\!&=&\!d\,(-4H\ddot H+2\9H^2-4dH^2\9H+(3\!-\!d)H^4)\nn\\
 H_{ij}\!&=&\!a^2\de_{ij}\,(4\dH+8dH\ddot H+6d\9H^2+4(d^2\!+\!d\!-\!3)H^2\9H
    +d(d\!-\!3)H^4)\nn\\
 ^{(1)}\!H_{00}\!&=&\!d^2(-4H\ddot H+2\9H^2-4dH^2\9H
    -{\ts\2}(d\!+\!1)(d\!-\!3)H^4)\nn\\
 ^{(1)}\!H_{ij}\!&=&\!a^2\de_{ij}\,d(4\dH+8dH\ddot H+6d\9H^2
    +(6d^2\!-\!4d\!-\!6)H^2\9H+{\ts\2}d(d\!+\!1)(d\!-\!3)H^4)\nn\\
 ^{(2)}\!H_{00}\!&=&\!d(-(d\!+\!1)H\ddot H+{\ts\2}(d\!+\!1)\9H^2
    -d(d\!+\!1)H^2\9H+{\ts\2}d(3\!-\!d)H^4)\nn\\
 ^{(2)}\!H_{ij}\!&=&\!a^2\de_{ij}\,((d\!+\!1)\dH+2d(d\!+\!1)H\ddot H
    +{\ts{3\02}}d(d\!+\!1)\9H^2+d(d^2\!+\!3d\!-\!6)H^2\9H\nn\\
    &&\!\hphantom{a^2\de_{ij}\,(}+\;{\ts\2}d^2(d\!-\!3)H^4) \label{Bsix}\eea
The Weyl tensor vanishes for the conformal flat FRW-metric, so that the
relation
$d(d\!-\!1)\,H_{\mu\nu}+2\,{}^{(1)}\!H_{\mu\nu}-4d\,{}^{(2)}\!H_{\mu\nu}=0$
results.
In $d\!+\!1=4$ spacetime dimensions the Euler number
$n=\int\!d^4\!x\,|g|^{1/2}(R_{\mu\nu\r\s}R^{\mu\nu\r\s}+R^2
-4\,R_{\mu\nu}R^{\mu\nu})$ is a topological invariant, hence its variational
derivative $H_{\mu\nu}+{}^{(1)}\!H_{\mu\nu}-4\,{}^{(2)}\!H_{\mu\nu}$
vanishes. The result (\ref{Bsix}) has been checked explicitly against these two
identities.\\
For the de~Sitter spacetime we have $\9H=0$ and all $H$-tensors
are vanishing for $d=3$.

\expandafter\ifx\csname ok\endcsname\relax
   \end{document}\fi
\clearpage
\expandafter\ifx\csname qd\endcsname\relax
   \documentstyle[12pt,twoside]{article}\let\qd=\quad \let\qqd=\qquad \def\qqqd{\qquad\qquad}

\let\a=\alpha \let\be=\beta \let\g=\gamma \let\de=\delta
\let\e=\varepsilon \let\z=\zeta \let\h=\eta \let\th=\theta
\let\dh=\vartheta \let\k=\kappa \let\la=\lambda
\let\r=\rho \let\s=\sigma
\let\om=\omega
\let\ph=\varphi
\let\Om=\Omega \let\Si=\Sigma \let\Th=\Theta
\let\La=\Lambda \let\G=\Gamma \let\D=\Delta

\def\0{\over } \def\1{\vec } \def\2{{1\over2}} \def\4{{1\over4}}
\def\5{\bar } \def\6{\partial } \def\7#1{{#1}\llap{/}}
\def\8{\tilde } \def\9{\dot }

\def\ul#1{$\underline{\hbox{#1}}$} \def\llp{\hbox to 0pt{\hss /\hskip1.5pt}}
\def\so{\supset\hbox to 0pt{\hss $\displaystyle -$\hskip1pt}}

\def\ts{\textstyle } \def\ds{\displaystyle }
\def\<{\langle } \def\>{\rangle } \def\lb{\left} \def\rb{\right}
\let\lra=\leftrightarrow \let\LRA=\Leftrightarrow
\let\then=\Rightarrow \def\triang{{\scriptstyle \Delta}}
\def\trian{{\scriptscriptstyle \Delta}}
\def\re{{\rm Re}} \def\im{{\rm Im}} \def\ord{{\cal O}}

\let\ap=\approx \let\eq=\equiv \let\ex=\times \let\hc=\dagger
      \let\ti=\tilde \let\bl=\biggl \let\br=\biggr
\let\bi=\choose \let\at=\atop \let\mat=\pmatrix

\let\nn=\nonumber
\def\bea{\begin{eqnarray}} \def\eea{\end{eqnarray}}
\def\bean{\begin{eqnarray*}} \def\eean{\end{eqnarray*}}
\def\beq{\begin{equation}} \def\eeq{\end{equation}}
\def\barr{\begin{array}} \def\earr{\end{array}}

\newcommand{\deflabel}[1]{#1\hfil}%
\newenvironment{deflist}[1]%
{\begin{list}{}%
{\settowidth{\labelwidth}{#1}%
\setlength{\leftmargin}{\labelwidth}\addtolength{\leftmargin}{\labelsep}%
\renewcommand{\makelabel}{\deflabel}}}%
{\end{list}}%
\fi
\expandafter\ifx\csname ok\endcsname\relax
   \addtolength{\topmargin}{-48pt}
   \addtolength{\textheight}{90pt}
   \addtolength{\evensidemargin}{-31pt}
   \addtolength{\oddsidemargin}{-12pt}
   \addtolength{\textwidth}{30pt}
   \addtolength{\footskip}{21pt}
   \setlength{\parindent}{0pt}\frenchspacing
   \begin{document}
\fi

\label{LastPage}

\expandafter\ifx\csname ok\endcsname\relax
   \end{document}\fi
\clearpage
\end{document}